\documentclass[]{jfm}

\usepackage{graphicx}
\usepackage{newtxtext}
\usepackage{newtxmath}
\usepackage[permil]{overpic}
\usepackage{natbib}
\usepackage{hyperref}
\usepackage{booktabs}
\usepackage[ruled,linesnumbered]{algorithm2e}
\usepackage{algorithmic}
\usepackage{amsmath}

\hypersetup{
    colorlinks = true,
    urlcolor   = blue,
    citecolor  = blue,
}

\newcommand{\RomanNumeralCaps}[1]
\linenumbers

\graphicspath{{images/}}

\usepackage[normalem]{ulem}
\usepackage{color}
\title{Prediction of flow and elastic stresses in a viscoelastic turbulent channel flow using convolutional neural networks}

\author{Arivazhagan~G.~Balasubramanian\aff{1,2}\corresp{\email{argb@mech.kth.se}},
Ricardo~Vinuesa\aff{1,2} \and Outi~Tammisola\aff{1,2}
}

\affiliation{
\aff{1}FLOW, Engineering Mechanics, KTH Royal Institute of Technology, Stockholm, Sweden
\aff{2}Swedish e-Science Research Centre (SeRC), Stockholm, Sweden
}

\begin{document}
\maketitle

\begin{abstract}
Neural-network models have been employed to predict the instantaneous flow close to the wall in a viscoelastic turbulent channel flow. Numerical simulation data at the wall is utilized to predict the instantaneous velocity-fluctuations and polymeric-stress-fluctuations at three different wall-normal positions in the buffer region. The ability of non-intrusive predictions has not been previously investigated in non-Newtonian turbulence. Our analysis shows that velocity-fluctuations are predicted well from wall measurements in viscoelastic turbulence. The models exhibit enhanced accuracy in predicting quantities of interest during the hibernation intervals, facilitating a deeper understanding of the underlying physics during low-drag events. The neural-network models also demonstrate a reasonably good accuracy in predicting polymeric-shear stress and the trace of the polymer stress at a given wall-normal location. This method could be used in flow control or when only wall information is available from experiments (for example, in opaque fluids). More importantly, only velocity and pressure information can be measured experimentally, while polymeric elongation and orientation cannot be directly measured despite their importance for turbulent dynamics. We therefore study the possibility to reconstruct the polymeric-stress fields from velocity or pressure measurements in viscoelastic turbulent flows. The results are promising but also underline that a lack of small scales in the input velocity fields can alter the rate of energy transfer from flow to polymers, affecting the prediction of the polymer-stress fluctuations. The present approach not only aids in extracting polymeric-stress information but also gives information about the link between polymeric-stress and velocity fields in viscoelastic turbulence.
\end{abstract}


\section{Introduction}

Viscoelastic fluids are widely used in industrial processes, and an understanding of complex-fluid behaviour becomes crucial for enterprises working with non-Newtonian flows. Further, in real-world scenarios, turbulent flows predominate, exhibiting chaotic and multi-scale dynamics. The turbulent flows of purely viscoelastic fluids have important technological implications due to increased mixing efficiencies at low Reynolds numbers and have also piqued the interest in drag-reduction and flow control communities at high Reynolds numbers. The addition of a tiny amount of polymer (parts per million) has proven efficient in reducing friction drag in pipe flows~\citep{virk71}, leading to substantial energy savings in fluid-transport applications. Such observation sparked intense research into the interaction between flow dynamics and polymers, in the context of skin-friction reduction.~\cite{lumley1969} postulated that the drag reduction attributed to polymer molecules arises from the extension of polymers, thereby increasing the effective extensional viscosity of the solution. This increase in extensional viscosity leads to the damping of small eddies, resulting in increased buffer-layer thickness. However,~\cite{de1986}~attributed the drag reduction to elastic effects rather than the viscous properties of polymers. This theory assumes that the turbulent kinetic energy~(TKE) is stored as elastic energy by the polymers, thereby influencing the energy cascade. Nevertheless, neither theory offers a comprehensive description of polymer induced drag reduction.

Various experimental investigations in channel and pipe flows~\citep{pinho1990,wei1992,den1997,warholic1999,warholic2001,ptasinski2001,ptasinski2003} have shown that polymers induce changes in the turbulence structure rather than simply attenuating it. These changes are characterized by an increase in streamwise turbulence intensity alongside a decrease in wall-normal turbulence. With access to direct numerical simulation (DNS) techniques~\citep{sureshkumar1997}, a deeper insight into the intricate interaction between polymers and flow fields has been probed. Notably,~\cite{dubief2004,dubief2005} pointed out that the polymers exhibited a structured energy exchange with the flow, primarily occurring around near-wall vortices. This increased polymer activity in the buffer region suggests that polymers are entrained around the vortices, resulting in the torque due to polymer stress opposing the rotation of the streamwise vortices~\citep{kim2007}, thereby extracting energy from near-wall vortices. Consequently, polymer stretching weakens the near-wall coherent structures, leading to a reduction of skin friction~\citep{stone2002,dubief2004,li2007,kim2007,kim2008}. Besides damping near-wall vortices, polymers also enhance the streamwise kinetic energy in near-wall streaks, thereby the net balance of these gives rise to a self-sustained drag-reduced turbulent flow~\citep{dubief2004,dubief2005}. Additionally, \cite{graham2010,graham2012} suggested that the turbulent flow is characterized by an alternating sequence of active and hibernating phases. These phases are distinguished by flow structures exhibiting strong vortices and wavy streaks during the active phase, and weak streamwise vortices during the hibernation phase, with viscoelastic flows characterized by increased hibernation intervals. Additional insights into the influence of polymer additives on drag reduction are detailed in~\cite{xi2019}.

Overall, the studies indicate that drag reduction due to polymer molecules results from the dynamic interaction between polymer molecules and turbulence. Nonetheless, the study of polymer drag reduction is also limited by the fact that the exact behaviour of polymers in turbulent flows is not fully deciphered. Moreover, the intermittent dynamics of polymers near the wall are predominantly explored through numerical simulations rather than experimental investigation of polymer stress. Consequently, there exists a scope to gain insights into polymer stress through potential experimental measurements of flow fields. However, experimental investigations of drag reduction in viscoelastic flows encounter limitations stemming from near-wall measurements and the capabilities of experimental techniques to accurately quantify the flow, without perturbing it. While a complete description of viscoelastic turbulence would require characterization of both velocity and polymeric stresses, accessing such polymer deformation directly from experimental measurements remains a challenging goal~\citep{stone2003}.

%Identifying such effects of elasticity sparked an interest to detect and understand drag-reducing behaviour in the presence of both elasticity and plasticity in the fluid. \cite{le}~analysed the simulation data of~\cite{rosti} using high-order dynamic mode decomposition, and compared the modes in complex fluids~(non-Newtonian fluids) with those in Newtonian fluids. Their results indicated that elasticity and plasticity of the complex fluids have similar effects on the coherent structures; in both cases, the flow is dominated by long streaks disrupted by rapid, localised perturbations. On the other hand, the Newtonian flow displays short streaks and more chaotic dynamics. \cite{daulet} found that the largest amount of drag reduction is achieved with a combination of finite elasticity and plasticity, and while the highly plastic flow (high Bingham numbers) relaminarizes, elasticity affects the relaminarization in a complex and non-monotonic fashion.

% Due to the multiple physical mechanisms that drive the flow, numerical simulation of such turbulent complex fluid flows becomes computationally expensive. Researchers have tried a variety of modeling strategies to circumvent the problems of simulating complex flows and to obtain an insight into the physics of the problem. The purpose of this research is to use machine learning approaches to enhance the modeling of complex fluid flows.

The phenomenon of drag reduction in viscoelastic turbulent flows has been explored from theoretical, experimental and numerical perspectives. However, in recent years machine-learning methods have expanded the possibilities in simulating~\citep{raissi20,eivazi22,vinuesa22}, predicting~\citep{guastoni21,eivazi21,yousif2023,yousif2023b} and controlling~\citep{vignon23,guastoni2023c} wall-bounded Newtonian turbulent flows~\citep{vinuesa24}. Data-driven methods also hold the potential for enhancing the understanding of the role of coherent structures in Newtonian ~\citep{cremades24} and viscoelastic turbulence dynamics~\citep{soleEVP}. In the domain of elastic stress predictions, researchers have explored the predictability of polymer stress components from velocity gradient using neural networks~\citep{nagamachi2019} and have also developed a framework for rheological model discovery from velocity field and information corresponding to boundary conditions and initial condition of polymer stress~\citep{ardekani}. %In the domain of viscoelastic turbulent flows, linear data-driven tool such as proper orthogonal decomposition~(POD) have been utilized to analyze velocity-fluctuations and changes in the inclination angle of coherent structures have been inferred~\cite{mohammadtabar2017}. 
These examples underscore the possibilities offered by data-driven methods to advance the understanding of polymers and flow field interaction in viscoelastic turbulent flow. However, a critical first step is to estimate the instantaneous polymer behaviour in an experimental setting. But, experimental studies encounter difficulties in accurately quantifying macromolecular extension, thereby significantly restricting the ability to fully characterize polymer behaviour~\citep{stone2003}. Moreover, near-wall measurements of flow fields in turbulent channel flows can also pose challenges. Nevertheless, non-linear machine-learning tools offer possibilities for improving experimental flow measurements and estimating flow and scalar quantities~\citep{vinuesa23,eivazi2024}. Neural-network models, and in particular fully-convolutional network (FCN) models have demonstrated excellent capabilities in predicting the instantaneous state of the flow using quantities measured at the wall in Newtonian turbulent flow~\citep{guastoni21}.

Hence in the present study, the idea of non-intrusive sensing has been applied to viscoelastic turbulent channel flow to predict the velocity fluctuations and polymeric-stress components near the wall using the quantities measured at the wall. To this end, FCN models are employed to predict the two-dimensional velocity fluctuations and polymeric-stress fluctuation fields at different wall-normal distances. The present work highlights the capability of a data-driven approach to model turbulence in complex-fluids flows. As a first step, directly providing a possible experimental measurement of wall-quantities or near-wall velocity fluctuations as inputs to the network model allows for an estimation of the polymer stress quantities in the near-wall region from experimental observations. Furthermore, the developed non-intrusive sensing models will also find useful applications in closed-loop control of wall-bounded turbulence in viscoelastic flows.

The paper is organised as follows. The methodology employed in the present work is introduced in~\S\ref{sec:methodology}. A description on the dataset to train the network models is provided in~\S\ref{subsec:dataset} and an overview of the network models can be found in~\S\ref{subsec:FCN}. A filter-based approach to identify the effects of small-scales on FCN predictions is detailed in~\S\ref{subsec:filtering}. In~\S\ref{sec:results}, the results obtained with different network models are discussed. The performance of the network models in different prediction types is outlined in~\S\ref{subsec:result_v}--\S\ref{subsec:result_psd}. The effects of filtering the small scales in the input velocity fluctuations on the predictions by FCN is discussed in~\S\ref{subsec:result_filter}--\S\ref{subsec:result_spec_analysis} and an interpretation of the predictions by FCN is provided in~\S\ref{subsec:result_interpretation}. Additional details regarding the error metrics for different types of predictions and the analysis of predictions of polymer shear stress can be respectively found in Appendix~\ref{app:A}--\ref{app:C}.

\section{Methodology}\label{sec:methodology}
\subsection{Dataset}\label{subsec:dataset}
In order to estimate the turbulent dynamics of the viscoelastic fluid from velocity fields, direct numerical simulation of viscoelastic turbulent channel flow is performed to generate the data necessary for modelling the relationship between near-wall velocity-fluctuations and polymer-stress-fluctuation quantities of interest using a fully-convolutional neural-network model. For the direct numerical simulation, a macroscopic continuum description of fluid-polymer interaction is considered. Specifically, a homogeneous dilute mixture of polymer in the solvent fluid is modelled with the \emph{finitely extensible non-linear elastic with Peterlin closure} (FENE-P) constitutive relation~(refer equation~(\ref{eqn:FENEP})). Although more sophisticated models~\citep{watanabe2014,shen2022} could be considered,~\cite{stone2003}~demonstrated that the FENE-P model can still yield reasonable predictions of the spatial distribution of stresses due to polymers and indicated that while the transient extension for the FENE-P model compares favourably to more detailed model such as bead-spring-chain models, it may overestimate elastic stresses in turbulent flow. Nonetheless, the FENE-P model can reproduce features of polymer drag reduction, including the maximum-drag reduction asymptote~\citep{graham2014}. Hence, in the present study, FENE-P model is utilized with the notion that it accurately describes the polymer physics of the considered viscoelastic liquid.

In the macroscopic description of the polymer model, we consider the statistics of end-to-end vector~$\mathbf{R}$ to characterize the average of the end-to-end vectors of a large number of polymer macro-molecules contained in a fluid particle. An essential statistical quantity in this description is the second-order correlation of the orientation vector~$\mathbf{R}$, known as the conformation tensor~$(\mathbf{A})$ which is obtained by:
\begin{equation}
    \mathbf{A} = \frac{A_{ij}}{R_0^2} = \frac{\left<R_i R_j\right>_f}{R_0^2}\,,
\end{equation}
where~$\left<\cdot\right>_f$ corresponds to the statistical average over a fluid particle and~$R_0^2$ corresponds to the root mean square~(RMS) of all the end-to-end vectors inside a fluid particle at rest.

The dimensional incompressible Navier--Stokes equations coupled with the evolution equation for the polymer conformation tensor are given by:

\begin{align}
     \label{seqn:mom}\frac{\partial \boldsymbol{u}}{\partial t} + \left(\boldsymbol{u} \cdot \nabla \boldsymbol{u}\right) &= -\frac{1}{\rho}\nabla p + \frac{\mu_s}{\rho} \nabla^2 \boldsymbol{u} + \left[ \frac{\mu_p}{\rho \lambda}\nabla \cdot \left( \mathcal{P}(\mathbf{A})\mathbf{A} \right) \right]\,,\\
     \nabla \cdot \mathbf{u} &= 0\,,\\
     \label{eqn:FENEP}\frac{\partial \mathbf{A}}{\partial t} + \boldsymbol{u}\cdot \nabla \mathbf{A} &= \mathbf{A}\cdot \nabla \boldsymbol{u} + \left(\nabla \boldsymbol{u}\right)^T\cdot \mathbf{A} - \frac{1}{\lambda}\left[ \mathcal{P}(\mathbf{A})\mathbf{A} - \mathbf{I} \right]\,,\\
     \mathcal{P}(\mathbf{A}) &= \frac{L_\mathrm{max}^2}{L_\mathrm{max}^2 - \mathrm{tr}(\mathbf{A})}\,,
\end{align}
where~$\boldsymbol{u}$ is the velocity with corresponding components in streamwise, wall-normal and spanwise directions represented by~$u,v,w$ and~$p$ is the pressure. In the remainder of the text, mean velocity components are denoted by~$\left<U\right>,\left<V\right>,\left<W\right>$ and the corresponding fluctuations by~$u,v,w$. In equation~(\ref{seqn:mom}),~$\rho$ is the density of the fluid with~$t$ denoting the time-scale of flow and~$\lambda$~denoting the relaxation time-scale of polymer stress. The solvent and polymer viscosity of the fluid are given by~$\mu_s,\,\mu_p$, respectively and they depend on the polymer concentration. The Peterlin function~$\mathcal{P}(\mathbf{A})$ accounts for the finite length of the polymer molecules with~$L_\mathrm{max}$ denoting the upper limit of the normalized polymer extension length, the point after which, the polymers cannot absorb more energy from the flow.

The polymer stress~$\boldsymbol{\tau_p}$ can be retrieved from the conformation tensor using the Kramer relationship:
\begin{equation}
    \boldsymbol{\tau_p} = \frac{\mu_p}{\lambda}\left(\mathbf{A}-\mathbf{I}\right)\,.
\end{equation}

Note that the mean of the polymer stress component is denoted by $\left<\tau_{p,\mathrm{ij}}\right>$ and the corresponding fluctuation by~$\tau_{p,\mathrm{ij}}$. One of the difficulties in simulating viscoelastic fluid flow comes from the preservation of positive definiteness of the conformation tensor~($\mathbf{A}$) during the evolution of turbulent flow at high Weissenberg number~($\text{\Wi}:=\lambda U_b/h$, with $U_b$ corresponding to the bulk velocity and~$h$ is the channel half-height; $\text{\Wi}$ quantifies the elastic forces with respect to the viscous forces, thereby indicating the degree of anisotropy in the flow).~\cite{sureshkumar1995} used artificial diffusion with the hyperbolic polymer-evolution equation. The artificial diffusion can be either local~\citep{dubief2004,dubief2005} or global~\citep{sureshkumar1997}. Despite the simplicity of the method, artificial diffusion affects the dynamics of the polymers at the small scales~\citep{nguyen2016,vincenzi2023}. Therefore, in the present study, the log-conformation approach~\citep{fattal2004} is utilized to ensure the positive-definiteness of the conformation tensor and thereby circumvent the high Weissenberg number problem.

The dataset for training and evaluation of the network model is obtained through a DNS of turbulent channel flow of viscoelastic fluid at a Reynolds number based on the bulk velocity of $\text{\Rey} = U_{b} h/\nu = 2800$ ($\nu (= (\mu_s+\mu_p)/\rho)$ denotes the total kinematic viscosity of the fluid), which corresponds to a friction Reynolds number $\text{\Rey}_{\tau} = 180$ (where $\text{\Rey}_{\tau}$ is defined in terms of $h$ and friction velocity $u_{\tau}$) for a Newtonian fluid. In this study, the turbulent channel flow simulations are performed at a Weissenberg number $\text{\Wi}=8$. The ratio of polymeric viscosity $\left(\mu_p\right)$ to the total viscosity~$\left(\mu_{s}+\mu_{p}\right)$, which is denoted by~$\alpha$, is set to $0.1$. The maximum polymer extensibility is set to $L=60$. The difficulties associated with proper rheological characterization of real fluids by adequate constitutive equations is an important area of research on its own and rather, we assume that the adopted model adequately describes the intended fluid properties.

\begin{figure}
\begin{centering}
\includegraphics[width=1\linewidth]{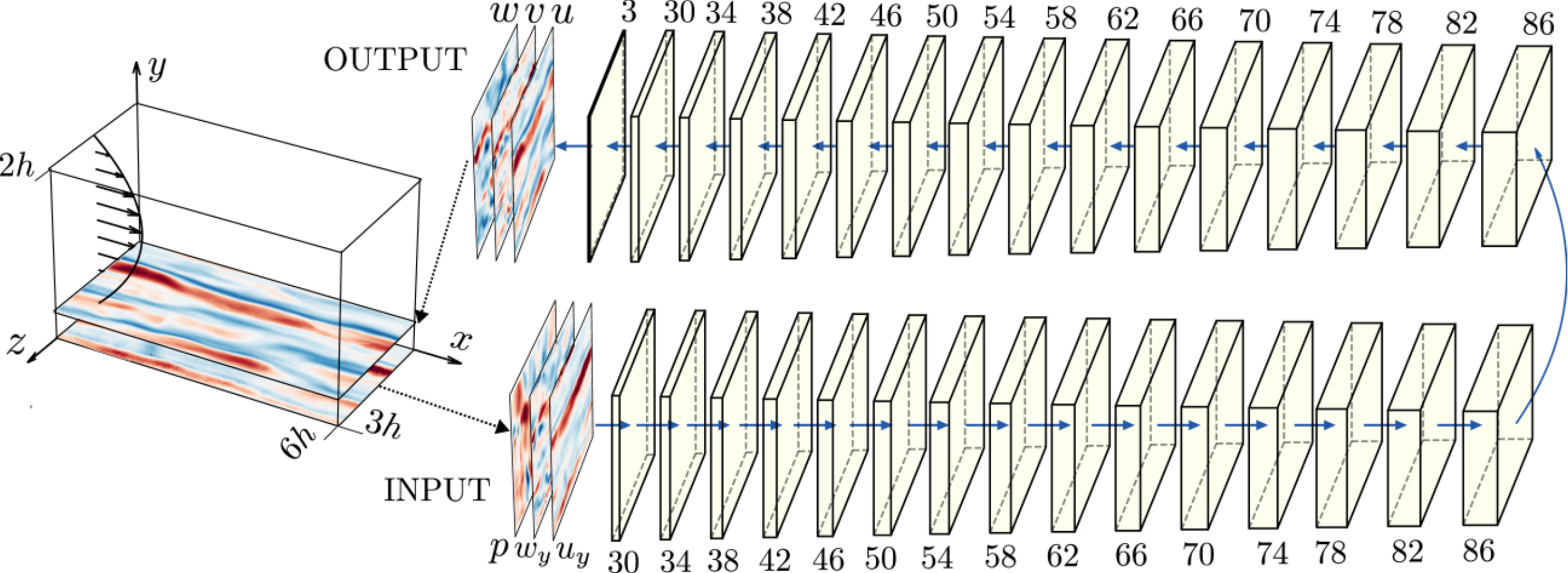}
\captionof{figure}{Typical workflow representation of V-prediction using fully-convolutional network (FCN) model. (Left) The computational domain for the channel flow and (right) FCN model with the corresponding number of kernels in each hidden layer is indicated.}
\label{fig:1}
\end{centering}
\end{figure}

\begin{figure}
\begin{centering}
\includegraphics[width=0.9\textwidth]{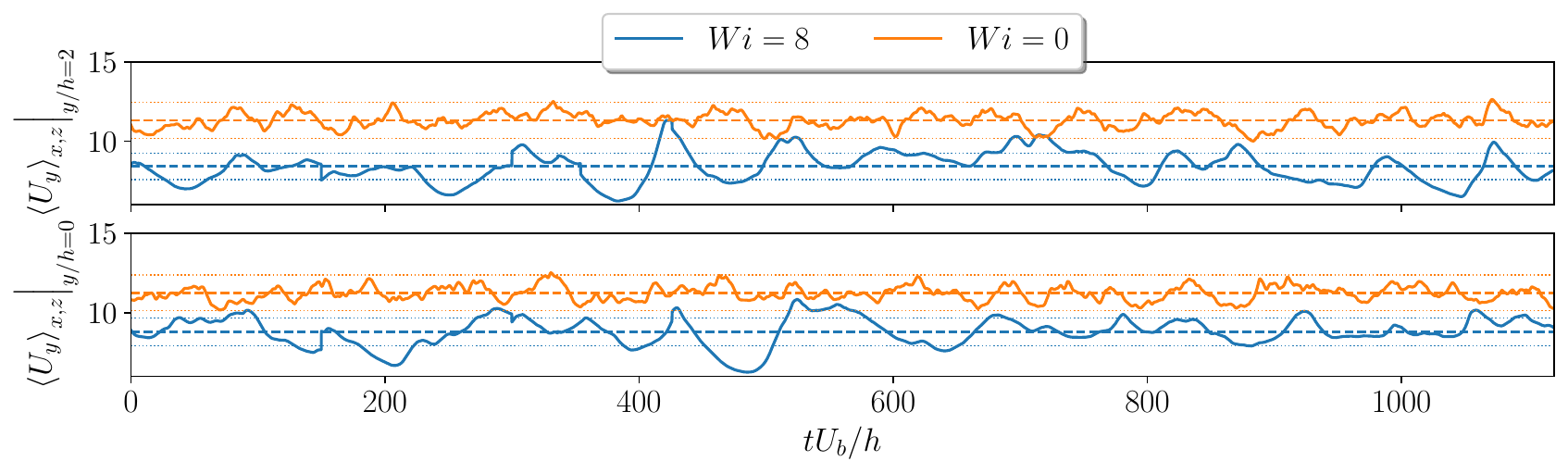}
\captionof{figure}{Time evolution of the wall-shear-rate in a viscoelastic channel flow corresponding to~$\text{\Wi}=8$ and Newtonian channel flow~$(\text{\Wi}=0)$ at (top)~$y/h=2$ and (bottom)~$y/h=0$. The dashed lines indicate the temporal mean and the dotted lines indicate the 10\% deviation from the temporal mean.}
\label{fig:2}
\end{centering}
\end{figure}

The viscoelastic turbulent channel flow is simulated using a finite-difference-based in-house code on a computational domain of size $6h\times 2h\times 3h$ in the streamwise~$\left(x\right)$, wall-normal~$\left(y\right)$ and spanwise directions~$\left(z\right)$, respectively as shown in figure~\ref{fig:1}. The readers are referred to~\cite{daulet} for a complete description of the viscoelastic turbulent channel simulation employed in this study. The computational domain is uniformly discretized using $1728 \times 576 \times 864$ grid points along $x,y$ and $z$, respectively. A spatial resolution of~$\Delta x^+ = \Delta y^+ = \Delta z^+ < 0.6$ is chosen to fully resolve the turbulent scales in the viscoelastic turbulent flow~\citep{rosti}. Here, the superscript `+' denotes the scaling in terms of the friction velocity $u_{\tau}$~($=\sqrt{\tau_w/\rho}$, where $\tau_w$ corresponds to the wall-shear stress) and viscous length $\ell^{*} \left(= \nu/u_{\tau}\right)$. Note that the value of $u_{\tau}$ obtained with $\text{\Wi}=8$ is lower than that in the Newtonian case~$\left(u_{\tau} \approx 180/\text{\Rey}_b;\,\text{since, }\text{\Rey}_{\tau,\,\text{\Wi}=0} \approx 180\right)$ %since $\text{\Rey}_{\tau,\,\text{\Wi}=8} \approx 0.87 \text{\Rey}_{\tau,\text{\Wi}=0}$ 
due to skin-friction reduction. Variation of the averaged wall-shear rate~$\left(\left<U_y\right>_{x,z} \rvert_{\mathrm{wall}}\right)$ obtained with~$\text{\Wi}=8$ is compared against the Newtonian case~$(\text{\Wi}=0)$ in figure~\ref{fig:2}. Here, $U_y$ corresponds to the wall-normal derivative of the streamwise velocity and $\left<\cdot\right>_{x,z}$ denotes the spatial averaging in $x$ and $z$ directions in the channel. From figure~\ref{fig:2}, identifying the hibernation intervals~(regions with low wall-shear stress) using area-averaged wall-shear rate as performed in~\cite{graham2010} with a threshold corresponding to $10\%$ of the mean shear-rate, we observe the presence of such low-drag events at~$\text{\Wi}=8$.  Note that the choice of threshold is arbitrary here and a definitive choice of the threshold value is absent in the literature. 
%At~$\text{\Wi} = 0$, the presence of `\emph{quiescent periods}' as described by~\cite{jiminez1991} is not directly observable but is hinted at around $tU_b/h \sim 500,700$ for $y/h = 2$ and at $tU_b/h \sim 860$ for $y/h = 0$. 
Effectively, for the considered viscoelastic turbulent flow at~$\text{\Wi}=8$, we observe a drag reduction of roughly~$20\%$ for the set of considered parameters in this study. From figure~\ref{fig:2}, it is evident that the fields at the wall~(which are provided as inputs to FCN, see~figure~\ref{fig:1}, \S\ref{subsec:FCN}) significantly deviate from the statistical mean for a considerable fraction of the total time. Thus, in this work, we aim to build a neural-network model that can predict viscoelastic turbulence quantities of interest, not only in the mean-flow but also in extreme wall-shear events with particular interest in hibernation intervals.
%It is important to note that only for viscous scaling in this study, the velocity scale considered is $u_{\tau} \approx \text{\Rey}_{\tau,\,\text{\Wi}=0}/\text{\Rey}_b$, which is the case for Newtonian fluid~$\left(\text{\Wi}=0\right)$. 

To this end, a database consisting of instantaneous fluctuation fields of wall-shear-stress components and wall-pressure fluctuation, as well as the two-dimensional velocity-fluctuation and polymeric-stress-fluctuation fields obtained at different wall-normal locations, $y^{+}= 13.6, 26.7$ and $44.2$~($y/h \approx 0.09, 0.17$ and $0.28$, respectively) is generated. Note that these wall-normal locations correspond respectively to $y^+ = 15, 30$ and $50$ for a Newtonian turbulent channel flow and hence, for simplicity, we refer to these locations as~$y^+ \approx 15, 30$ and $50$, respectively in this study. Buffer region is probed because of its importance in terms of production and dissipation of turbulent energy and is significantly modified in drag-reducing flows with polymer additives~\citep{den1997,stone2003}.

The simulations are run for $\sim 120 h/u_{\tau}$ time-units and a total of 40,600 samples is obtained with a sampling period of $\Delta t_s^+ \approx 1$ for training the network model. The sampled instantaneous two-dimensional fields are down-sampled to a resolution of $(N_x \times N_z =)\; 432 \times 432$ in $x$ and $z$, respectively. The fields on both walls are utilized in this study, and they are split into training and validation set with a ratio of 4 to 1. The computational resources required to carry out the simulation and generate the necessary training dataset amounted to approximately 1.5 million core-hours. 

The network models are evaluated with the samples in the test dataset which consists of 10,000 samples. The samples in the test dataset are chosen from a time-interval (in the sampled time series) that corresponds to at least 60 flow-through times apart from the samples in the training dataset to ensure minimal auto-correlation between the samples in the training and test dataset. The temporal length for the number of samples considered in the test dataset corresponds to $\approx 50 h/u_{\tau}$ time units, whereas a temporal averaging over 10--15 eddy-turnover times is sufficient to obtain good convergence of turbulence statistics~\citep{li2006}.

\subsection{Neural-network model} \label{subsec:FCN}
%We utilize a specific type of neural network, the convolutional neural network in this work because of its success in computer vision~\cite{le_cun}. The CNN consists of convolutional layers that are defined in terms of kernels (or filters) which convolves with the inputs and produces a transformed output. The learnable parameters are contained in the kernels and the transformed output is called the feature map which contains certain features extracted from the input image. Multiple feature maps are stacked and followed by an activation function to obtain a non-linear transformation. The activated feature maps are sequentially fed into the successive convolutional layers as input, which helps in combining extracted features to predict larger and more complex features with a deeper CNN. 
% In this research, we are using existing architectures. The above results may be further improved by the latest neural networks, but what is essentially important is not simply the improvement of estimation accuracy. In the case of the surrogate-model construction, inference that does not violate mathematical realizability and the physics laws should be guaranteed. In the case of image recognition, the identification of features is a crucial issue. For this purpose, PINN (Physics-Informed Neural Network) and GAN (Generative Adversarial Network) must be effective. 
In this work, a fully convolutional neural network similar to the one proposed by~\cite{guastoni21} is used, with an increased number of hidden layers~(see figure~\ref{fig:1}) to obtain a more complex combination of abstract turbulent features identified by the kernels in the network. Here, we utilize an existing architecture, acknowledging that further enhancements could be achieved with newer architectures that require extensive datasets. Our focus is on proposing a methodology for viscoelastic stress predictions in turbulent flows aimed towards experimental applications and in establishing baseline performance using current convolutional architectures. The considered FCN consists of 30 hidden layers with a total number of trainable parameters amounting to $985,105$. The convolution operations are followed by batch normalization and a rectified-linear-unit (ReLU) activation function. The inputs to the network are normalized using respective mean and standard deviation of the fields from the training dataset and the outputs are normalized using the corresponding standard deviation values. The choice of loss function~($\mathcal{L}$) in the network is the mean-squared error (MSE) between the instantaneous predicted and DNS fields:
\begin{align}
\label{eqn_loss_fn}
    \mathcal{L}(\boldsymbol{u}_{\mathrm{FCN}};\boldsymbol{u}_{\mathrm{DNS}}) &=
\frac{1}{N_{x}N_{z}} \sum_{i=1}^{N_{x}} \sum_{j=1}^{N_{z}} | \boldsymbol{u}_{\mathrm{FCN}}(i,j) - \boldsymbol{u}_{\mathrm{DNS}}(i,j) |^{2}\,,
\end{align}
which helps the network to learn the large-scale features first and then progressively optimize the trainable parameters to minimize the errors at finer scales~\citep{xu2019}. The subscripts `$\cdot_\mathrm{DNS}$' and `$\cdot_\mathrm{FCN}$' are respectively used to denote the DNS samples and corresponding predictions from FCN. Each network model is trained using $4 \times \mathrm{A100}$ graphics-processing units (GPUs), amounting to approximately 3,500 GPU-hours for training a single network model.

In this study, three types of predictions have been undertaken to highlight the capability of FCN models to reconstruct the near-wall visco-elastic turbulence fields. In V-predictions (indicating velocity predictions), the streamwise wall-shear, spanwise wall-shear and pressure field at the wall are utilised to predict the streamwise, spanwise and wall-normal velocity-fluctuations. This allows us to assess whether velocity fields can also be predicted in viscoelastic turbulence exhibiting periods of hibernation. The performance of neural-network models in predicting the fluctuations of polymeric-shear stress ($\tau_{p,\,\mathrm{xy}}$) and fluctuations of trace of polymer stress ($\mathrm{tr}(\tau_{p})$) at a given wall-normal location using the true velocity-fluctuation fields at the same location are denoted as E-predictions~(signifying prediction of elastic stress quantities of interest). Finally in V-E-predictions, the FCN model is used to predict the fluctuations of polymeric-shear stress and diagonal components of polymer-stress tensor at a target wall-normal distance directly from wall inputs, with auxiliary predictions of corresponding velocity-fluctuations at the considered wall-normal location~\footnote{The auxiliary predictions of velocity-fluctuations at a wall-normal location is utilized in V-E predictions to obtain an increase in the accuracy of prediction of polymeric-stress quantities.}. It is worth noting that the mean of polymer shear stress can be obtained either from experimentally quantified stress-deficit~\citep{warholic1999} or from numerical simulations. Additionally, the FCN model performs well in predicting the mean of polymer-stress quantities, which stems from the use of the mean-squared error as a loss function that tends to regress to the mean and hence the relative errors in predicting the mean polymer stress quantities from FCN are lower than~$2\%$. Consequently, our primary focus in this study is on retrieving the instantaneous fluctuations of polymer-stress using corresponding fluctuations of input quantities.

The mean-absolute error between the predictions and DNS fields~(denoted by~$\mathrm{MAE}$) is reported for different types of predictions, which is calculated as
\begin{align}
\label{eqn_mae}
    \mathrm{MAE}(u) = \left<| u_{\mathrm{FCN}} - u_{\mathrm{DNS}} |\right>_{x,z}\,.
\end{align}
The error metrics described in this section is computed for each component of~$\boldsymbol{u}$ and for the polymer-stress quantities of interest~(refer Appendix~\ref{app:A}).
The network performance is also evaluated from a statistical point of view in terms of the relative error in predicting the corresponding root-mean-squared (RMS) quantities between the true~(DNS) fields from test dataset and the fields predicted by the FCN~(indicated by~$E_\mathrm{RMS}$), and is given by:
\begin{align}
\label{eqn_Erms}
E_{\mathrm{RMS}} (u)\;\left[\%\right] = 100 \cdot \frac{|u_{\mathrm{FCN,RMS}}-u_{\mathrm{DNS,RMS}}|}{u_{\mathrm{DNS,RMS}}}\,,
\end{align}
whereas the instantaneous correlation coefficient between the predicted and the DNS fields is defined as:
\begin{align}
\label{eqn_R}
R (u) = \frac{\left<u_{\mathrm{FCN}} u_{\mathrm{DNS}}\right>_{x,z,t}}{u_{\mathrm{FCN,RMS}} u_{\mathrm{DNS,RMS}}}\,,
\end{align}
with $\left<\cdot\right>_{x,z,t}$ corresponding to the average in space~($x,z$) and time~($t$; denotes the samples in the test dataset) and subscript~$\cdot_\mathrm{RMS}$~refers to root-mean-squared quantities.
%For type-I predictions, $E_\mathrm{RMS}$ and~$\mathrm{MAE}$ would correspond respectively to the artificial turbulent kinetic energy and artificial turbulence intensity in the predictions. 
Note that the performance metrics reported in this study are obtained from the mean of at least three different network models to include the effects of stochasticity introduced by the random initialization of kernel weights in FCN and random sampling of mini-batches during the training process. The instantaneous correlation coefficient between the predicted and DNS fields averaged over the samples in the test dataset, is also highlighted. To evaluate the distribution of energy across the various scales in the flow, a comparison of the pre-multiplied two-dimensional~(2D) power-spectral density~(PSD)~$k_z k_x \phi_{ij} \left(\lambda_x^+,\lambda_z^+\right)$ between DNS fields and the predictions is performed. Here, $\phi_{ij}$ is the power-spectral density obtained for the quantity `$ij$' and $k_x, k_z$ respectively denote the wavenumbers in streamwise and spanwise directions, with the corresponding wavelengths in viscous units denoted by~$\lambda_x^+$ and $\lambda_z^+$. 
%\begin{equation} \label{eq:loss}
%\mathcal{L}(\mathbf{\widehat{u}}_\mathrm{FCN};\mathbf{\widehat{u}}_\mathrm{DNS})=\frac{\sum_{i=1}^{N_{x,s}} \sum_{j=1}^{N_{z,s}} \left | \mathbf{\widehat{u}}_\mathrm{FCN}(i,j) - \mathbf{\widehat{u}}_\mathrm{DNS}(i,j)\right |^{2}}{N_{x,s} N_{z,s}},
%\end{equation}

\subsection{Low-pass filtering of velocity fluctuations}\label{subsec:filtering}
In order to identify the effects of small-scale features in the input velocity-fluctuation fields on the predictions of polymer stress~(see \S\ref{subsec:result_spec_analysis}--\S\ref{subsec:result_interpretation}) and thereby to recommend scale requirements for inputs from possible experimental investigations, we employ~\emph{low-pass} filtering of velocity-fluctuations with a threshold wavelength denoted by~$\lambda_c^+$. This threshold~$\lambda_c^+$ represents the smallest wavelength of the turbulent scales present in the input velocity-fluctuation fields to the FCN. For instance, in velocity-fluctuation fields sampled from DNS with $432\times 432$ data-points respectively in the streamwise and spanwise directions, the minimum wavelengths correspond to~$\lambda_x^+ = 5,\,\lambda_z^+ = 2.5$ and hence,~$\lambda_c^+ = 2.5$ for the DNS sampled fields. By filtering the sampled DNS velocity-fluctuation fields at a certain~$y^+$ to contain small-scales above~$\lambda_c^+$, we ascertain that~$\lambda_x^+ = \lambda_z^+ \ge \lambda_c^+ (> 2.5)$. However, the predictions of polymer stress quantities from the FCN aim to capture all the scales above 2.5 viscous-length units, regardless of the input~$\lambda_c^+$. In other words, the sampled DNS polymer-stress quantities serve as the true reference for the supervised training process of FCN for all the different network models with corresponding~$\lambda_c^+$ of input velocity-fluctuations.

The filtering is performed in the wavenumber space~$k_x,k_z$ and the corresponding procedure for the filtering process is outlined in~Algorithm~\ref{alg:filter}.

\begin{algorithm}[H]
%    \SetAlgoLined
    \DontPrintSemicolon
    \KwIn{$\lambda_c^+$, 2D field sequence~${\chi}_i \in \left\{u_1,u_2,...u_N,v_1,v_2,...v_N,w_1,w_2,...w_N\right\}$}
    \KwOut{2D filtered field sequence~${\zeta}_i \in \left\{\tilde{u}_1,\tilde{u}_2,...\tilde{u}_N,\tilde{v}_1,\tilde{v}_2,...\tilde{v}_N,\tilde{w}_1,\tilde{w}_2,...\tilde{w}_N\right\}$}    
    \For{$i \leftarrow 1$ \KwTo $3N$}{
    \hspace{1em} Compute 2D-FFT of ${\chi}_i$: $\mathcal{U}(k_x,k_z) \leftarrow \text{FFT}(\chi_i)$\;
    
    \hspace{1em} 2D-Shift FFT: $\mathcal{U}_\mathrm{shift}(k_x,k_z) \leftarrow \text{FFTshift}(\mathcal{U})$\;

    \hspace{1em} \textbf{if} {$\sqrt{\left(\frac{2\pi}{k_x \ell^*}\right)^2 + \left(\frac{2\pi}{k_z \ell^*}\right)^2} < \lambda_c^+$} \textbf{then}\;
    
    \hspace{2.5em} $\mathcal{U}_\mathrm{shift}(k_x,k_z) \leftarrow 0$\;

    \hspace{1em} \textbf{end}\;

    \hspace{1em} Inverse 2D-shift FFT: $\widetilde{\mathcal{U}}(k_x,k_z) \leftarrow \text{IFFTshift}(\mathcal{U}_\mathrm{shift})$\;

    \hspace{1em} Compute inverse 2D-FFT of~$\widetilde{\mathcal{U}}$ : $\zeta_i \leftarrow \text{IFFT}(\widetilde{\mathcal{U}})$\;
     }
    \Return{$\boldsymbol{\zeta}$}
    \caption{\label{alg:filter} Low-pass filtering of input velocity fields}
\end{algorithm}

The relative decrease in the turbulent kinetic energy in the input velocity-fluctuations sampled from DNS~(denoted by~$\mathcal{K}_{\mathrm{DNS}}$) and after the filtering process~(denoted by~$\mathcal{K}_{f}$) for a given~$\lambda_c^+$ is quantified as:
\begin{equation}
    \Delta \mathcal{K} = 100 \cdot \frac{\mathcal{K}_\mathrm{DNS}-\mathcal{K}_{f}}{\mathcal{K}_\mathrm{DNS}}\;[\%] = 100 \cdot \frac{\left<\boldsymbol{\chi}^2\right>-\left<\boldsymbol{\zeta}^2\right>}{\left< \boldsymbol{\chi}^2 \right>}\;[\%]\,,
\end{equation}
where $\left<\cdot \right>$ corresponds to the spatial and temporal averaging of data yielding a scalar output. Here,~$\boldsymbol{\chi}$ corresponds to the sequence of sampled velocity-fluctuation snapshots from DNS and~$\boldsymbol{\zeta}$ is the sequence of respective filtered fields obtained from filtering process.
\section{Results}\label{sec:results}
The predictions of the trained network models are compared with the data obtained from DNS. The performance is assessed from a qualitative point of view and subsequently from a quantitative aspect, based on predictions of instantaneous fields, turbulent statistics and the two-dimensional power spectral density. Further, the importance of small-scale velocity fluctuations to successfully retrieve the small-scale polymer stress fluctuations is highlighted.

\subsection{Prediction of velocity-fluctuations}\label{subsec:result_v}
\begin{figure}
\begin{centering}
\includegraphics[width=1\linewidth]{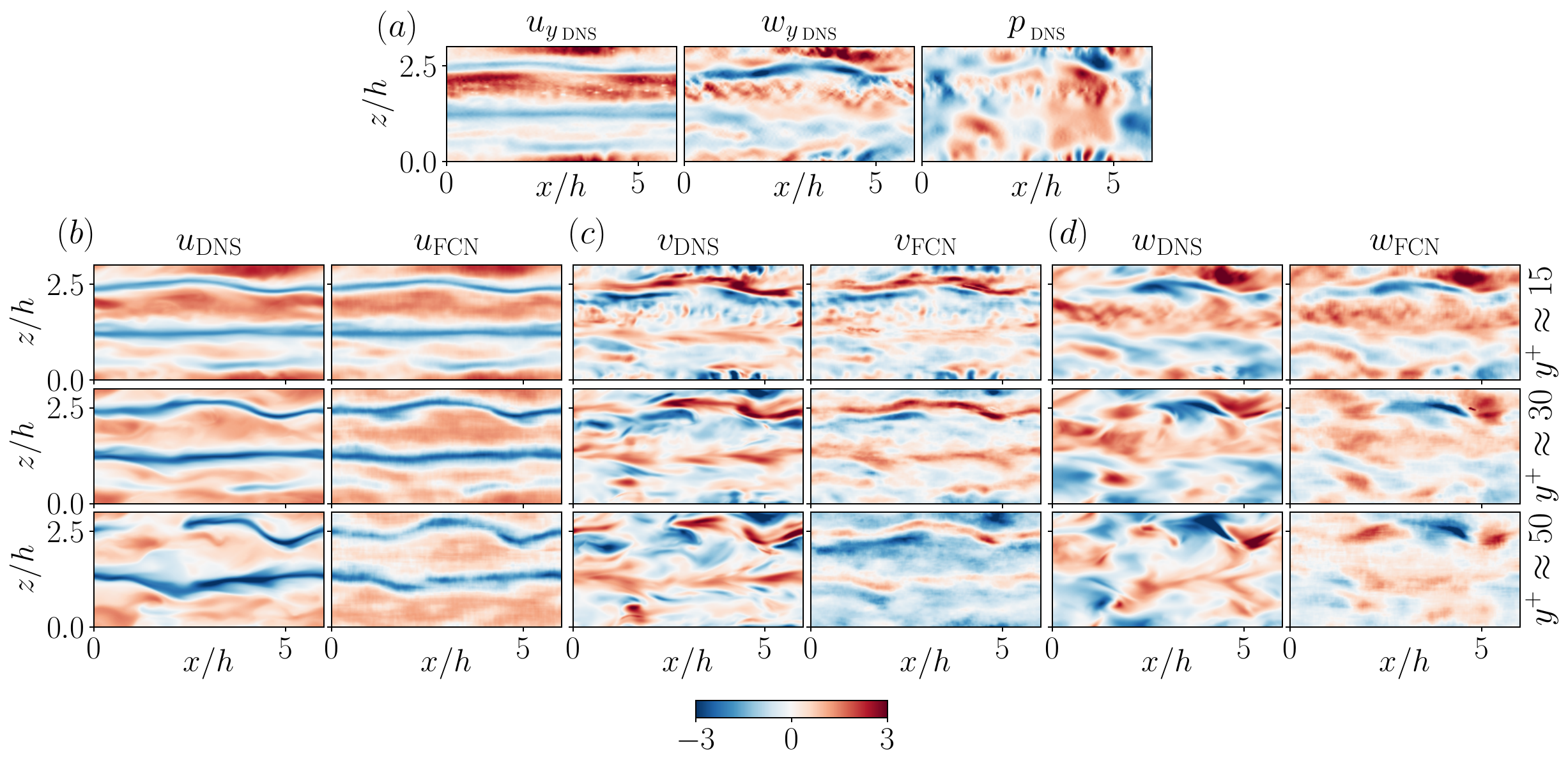}
\captionof{figure}{Sample instantaneous~$(a)$~normalized wall inputs to the FCN compared with the instantaneous velocity-fluctuation fields in the $(b)$ streamwise, $(c)$ wall-normal and $(d)$ spanwise directions, at different wall-normal positions. In $(b-d)$: (left) corresponds to the DNS field and (right) shows the corresponding V-predictions from FCN. The fields are scaled with the corresponding RMS values.}
\label{fig:velocity}
\end{centering}
\end{figure}

The predicted velocity-fluctuations at different target wall-normal positions using streamwise and spanwise wall-shear rate and wall pressure are qualitatively inspected. A sample prediction of the instantaneous velocity-fluctuation fields for the case of V-predictions is shown in figure \ref{fig:velocity} (corresponding to an instant in the test dataset where the input wall-shear rate is higher than the mean wall-shear rate). We note that the predicted velocity fields are visually well correlated with the DNS fields at different target wall-normal locations. The linear correlation coefficient between the predicted and true streamwise-velocity fluctuation fields exceeds 99\% for predictions at~$y^+ \approx 15$, and gradually declines but remains above 80\% at $y^+ \approx 50$. The RMS quantities of the streamwise velocity-fluctuation fields at $y^+ \approx 15,30,50$ are predicted with less than~$\left(E_\mathrm{RMS} < \right)$ 3\%, 6\% and 15\% error, respectively. With an increasing separation distance~(wall-normal distance between the wall fields and the target velocity-fluctuation fields), the fields are less-correlated and thereby the performance of the network also decreases. Because of this, the RMS-normalized mean-absolute errors in the predicted streamwise-velocity fluctuations are 0.14, 0.29 and 0.47 at~$y^+ \approx 15,30,50$, respectively~(see also figure~\ref{fig:intensity1}). The performance metrics for different network models are summarized in Appendix~\ref{app:A}.

The MAE in the wall-normal and spanwise fluctuation fields remained below~$0.025$ in the different target wall-normal locations considered in the study~(refer figure~\ref{fig:intensity1} for normalized quantities). However, the $E_\mathrm{RMS}$ values in the wall-normal and spanwise-velocity fluctuations are at least twice as large as those obtained in the RMS prediction of the streamwise component at the respective wall-normal locations. This is due to the influence of the polymers, which reduce turbulence by opposing the downwash and upwash flows generated by near-wall vortices~\citep{dubief2004,dubief2005}. Due to the absence of polymeric-stress information in the inputs to the network model for V-predictions, an accurate representation of the turbulence statistics in the spanwise and wall-normal fluctuation components becomes challenging.

\begin{figure}
\begin{centering}
\includegraphics[height=120pt]{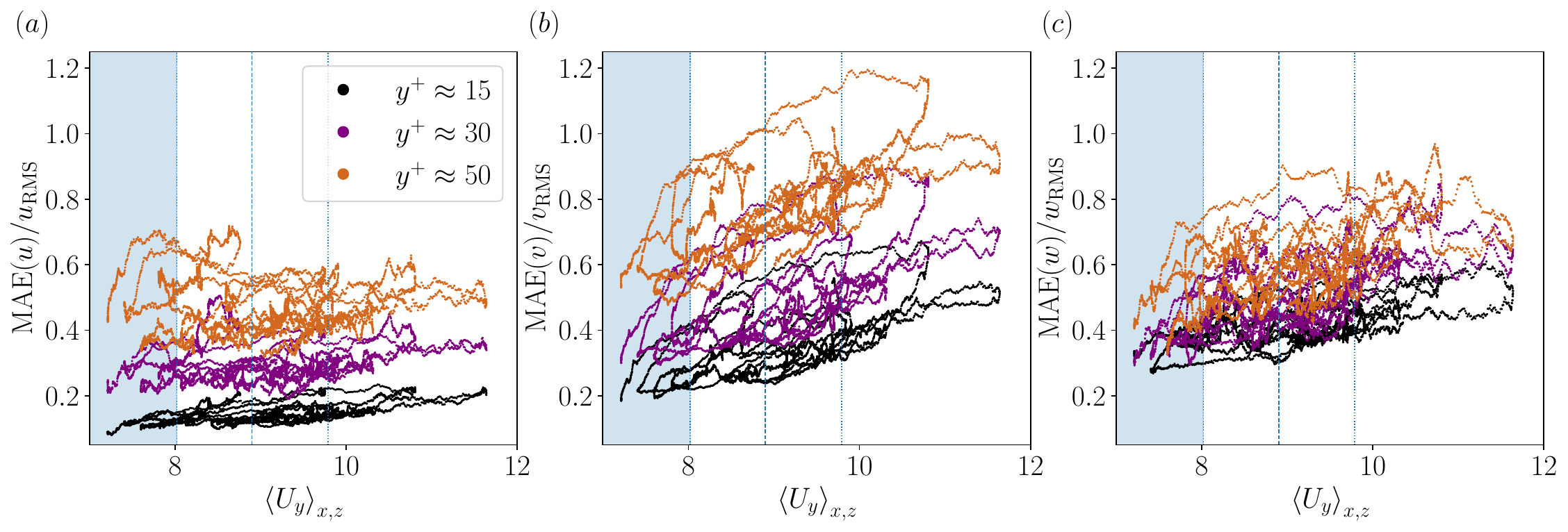}
\captionof{figure}{Variation of the RMS-normalized mean-absolute errors of~$(a)$~streamwise, $(b)$~wall-normal and~$(c)$~spanwise velocity components in V-predictions at different wall-normal locations with respect to the wall-shear rate. The markers correspond to the mean absolute error in the instantaneous sample for the test dataset. The shaded region corresponds to the hibernation interval identified with 90\% of~$\left<U_y\right>_{x,z,t}$. The dashed vertical lines indicate the temporal mean and the dotted vertical lines indicate the 10\% deviation from the temporal mean.}
\label{fig:intensity1}
\end{centering}
\end{figure}

It should be emphasized that the network model is explicitly optimized for predicting instantaneous fields rather than reproducing the turbulence statistics. This emphasis is rooted in the motivation for non-intrusive sensing in an experimental setting, aimed at understanding the near-wall dynamics of viscoelastic turbulent channel flow. In addition, optimizing network models to accurately replicate turbulence statistics obtained from DNS could lead the model to learn the mean-flow behaviour with a lower~$E_{\mathrm{RMS}}$. This may also entail a compromise, as predictions during hibernating intervals could potentially become less accurate.

When assessing the accuracy of the instantaneous predictions based on mean-absolute errors, as illustrated in figure~\ref{fig:intensity1}, it becomes apparent that the MAE (in each test sample) varies with wall-shear rate for different target wall-normal locations. Specifically, in instances of low-wall-shear-rate, the absolute errors are notably lower and increase with wall-shear rate. This is due to the fact that low-drag events typically exhibit reduced fluctuation intensity, which increases with wall-shear rate, leading to increased concentration of energy in small-scale features. Consequently, the network encounters relative difficulty in accurately predicting these small-scale features, resulting in higher prediction errors at large wall-shear-rate inputs.
It is worth noting that the variation of MAE (in each test sample) with wall-shear rate stems from the selection of the loss function utilized in the network. 
%If L1 norm were to be chosen as the loss function instead of MSE, the performance of instantaneous predictions could potentially become nearly independent of the input wall-shear rate.
Nevertheless, the obtained network model exhibits superior predictive performance in capturing velocity-fluctuation fields during low-wall-shear rate events. This observation underscores the potential utility of such models in obtaining sufficiently accurate velocity-fluctuations in an experimental setting, more particularly for studying hibernation events in detail.

\subsection{Prediction of polymer stress}\label{subsec:result_ve}
\begin{figure}
\begin{centering}
\includegraphics[width=1\linewidth]{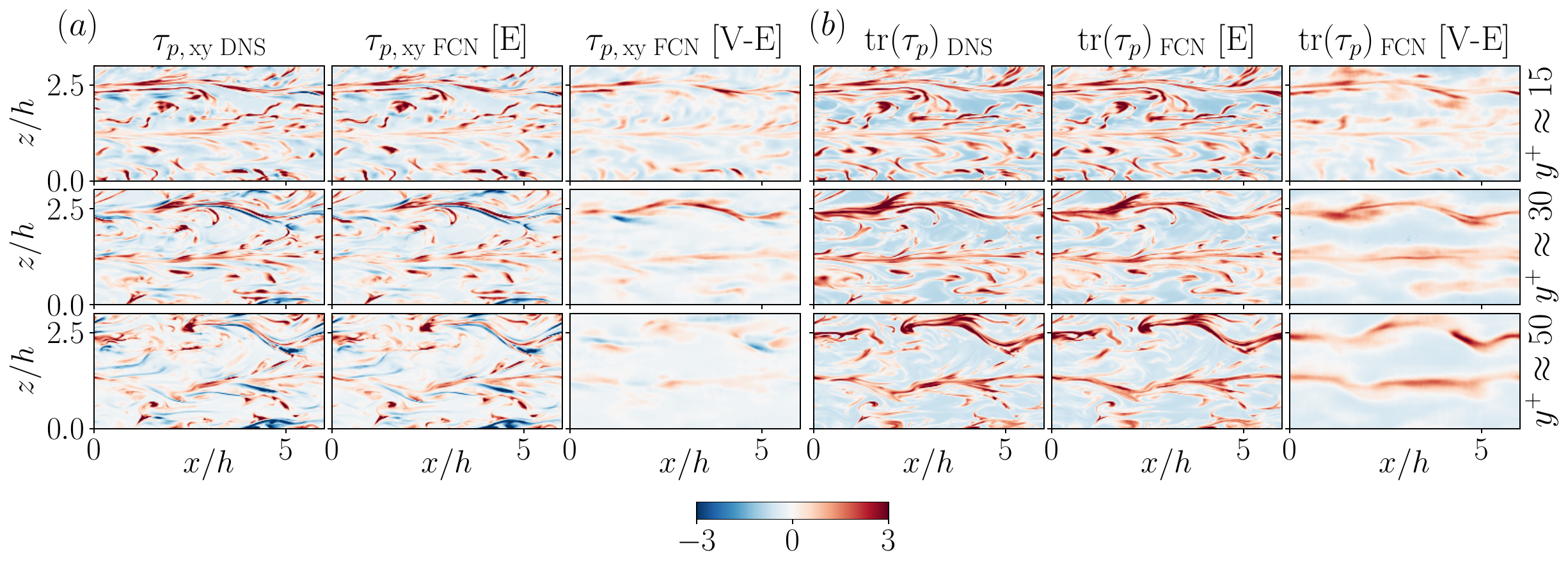}
\captionof{figure}{A sample fluctuation field corresponding to $(a)$ polymer-shear-stress and $(b)$ trace of the polymer-stress, at different-wall normal locations. In $(a-b)$: (left) corresponds to the DNS field, (middle) shows the E-predictions and (right) corresponds to V-E-predictions from FCN. The fields are scaled with the respective RMS values.}
\label{fig:stress}
\end{centering}
\end{figure}

The predicted velocity fluctuations at various target wall-normal positions show a good agreement with the reference DNS quantities. Following this, the focus is to estimate the polymer stress quantities of interest based on velocity fluctuations. Additionally, an effort is made to retrieve polymer stress using only wall information, which is advantageous in experimental settings. This section begins with a qualitative analysis of the predicted polymer stress fields. A sample predicted field of polymer stress components (corresponding to the same time instant as the wall inputs shown in figure~\ref{fig:velocity}$a$) for E-predictions and V-E-predictions is shown in figure \ref{fig:stress}. Overall, the large-scale features in the polymer-stress quantities of interest are visually in good agreement with DNS. For E-predictions, where polymeric-stresses are predicted from DNS velocity fields at the same location there is no separation distance between the input and target fields and the linear correlation coefficient between the predicted and DNS polymer-shear stress, as well as with the trace of the polymer stress remained more than 90\% for the different target wall-normal positions. Moreover, $E_\mathrm{RMS}$ remained below 15\% for E-predictions of~$\tau_{p,\mathrm{xy}}$ and $\mathrm{tr}(\tau_p)$ at different target wall-normal distances.

Note that in V-E-predictions the polymeric-stresses are predicted directly from the wall inputs, without having access to the true velocity fields at the wall-normal location where those stresses are predicted. Instead, predicted auxiliary velocity fields at that location (together with wall inputs) are used to predict the polymeric-stress fields. The obtained errors of around $40\%$ indicate that a small error in predicting velocity-fluctuation fields significantly impacts the errors in predicting the polymeric-stress fields~(see also~\S\ref{subsec:result_filter}), suggesting that the auxiliary velocity-fluctuation fields in V-E-predictions lack certain information that is connected to the polymeric activity in the small wavelengths. Nevertheless, the large-scale structures in the predicted polymeric-stress fields for V-E-predictions exhibit a qualitative agreement with the reference, as observed in figure~\ref{fig:stress}.

\begin{figure}
\begin{centering}
\includegraphics[width=1\linewidth]{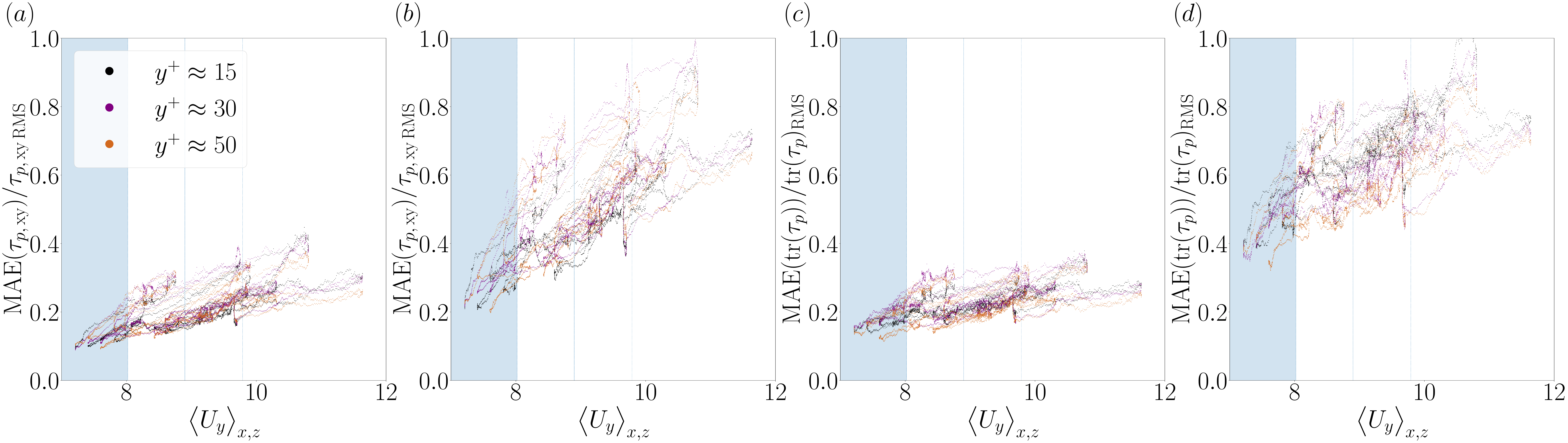}
\captionof{figure}{Variation of the RMS-normalized mean-absolute errors of polymer-shear stress in~$(a)$~E-predictions, $(b)$~V-E-predictions and trace of polymer stress in~$(c)$~E-predictions and $(d)$~V-E-predictions with respect to the wall-shear rate. The markers correspond to the mean absolute error in the instantaneous sample in the test dataset. Shaded regions correspond to the identified hibernation interval with 90\% of~$\left<U_y\right>_{x,z,t}$. The dashed vertical lines indicate the temporal mean and the dotted vertical lines indicate the 10\% deviation from the temporal mean.}
\label{fig:intensity23}
\end{centering}
\end{figure}

Examining the accuracy of the instantaneous predictions based on mean-absolute errors, as depicted in figure~\ref{fig:intensity23} for E-predictions and V-E-predictions, reveals a similar trend in MAE (in each test sample) with respect to wall-shear rate as observed in V-predictions. Overall, the absolute errors increase with wall-shear rate. Further, the magnitude of such absolute errors in the field is nearly doubled for V-E-predictions~(figure~\ref{fig:intensity23}$b$, $d$) compared to E-predictions~(figure~\ref{fig:intensity23}$a$, $c$). Moreover, the MAE in predicting polymer-stress quantities of interest remains relatively constant across various target wall-normal positions with respect to the corresponding RMS quantities for E-predictions and with marginal distinctions in V-E-predictions. %Since the RMS of~$\tau_{p,\,\mathrm{xy}}$ increases with wall-normal distance, the MAE in predicting~$\tau_{p,\,\mathrm{xy}}$ also increases with $y^+$.

\subsection{Energy distribution across different length scales}\label{subsec:result_psd}
The statistical analysis of the predicted velocity fluctuations using V-predictions and the predicted polymer stresses using E-predictions closely matches the results obtained from DNS. However, there is a significant deviation in the second-order statistics of polymer stress quantities predicted using only the wall-shear rate and wall pressure~(V-E predictions). To understand the reasons behind this, the distribution of energy across different length scales in the FCN predictions is examined and compared with the reference DNS.

\begin{figure}
\begin{centering}
\includegraphics[height=200pt]{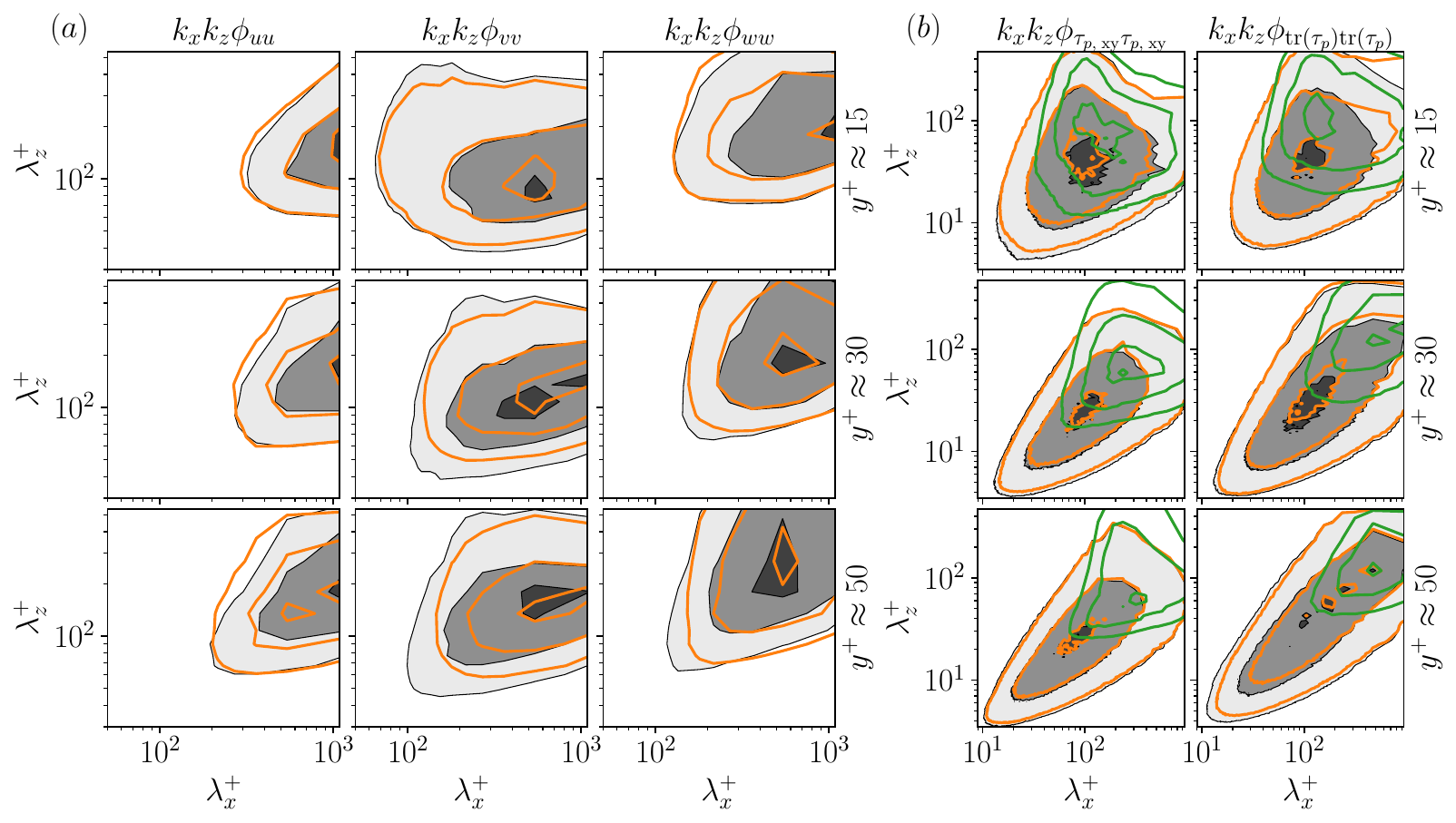}
\captionof{figure}{Pre-multiplied two-dimensional power-spectral densities of ($a$, left)~the streamwise, ($a$, center)~wall-normal, ($a$, right)~spanwise velocity components and ($b$, left)~polymer shear-stress, ($b$, right)~trace of polymer stress at~(top)~$y^+\approx 15$, (middle)~$y^+ \approx 30$ and (bottom)~$y^+ \approx 50$. The contour levels contain 10\%, 50\% and 80\% of the maximum power-spectral density. Shaded contours refer to DNS data, while contour lines correspond to ($a$)~V-predictions, ($b$, orange)~E-predictions and ($b$, green)~V-E-predictions.}
\label{fig:spectra_velocity}
\end{centering}
\end{figure}

The distribution of energy in the predicted and DNS data across different scales are compared through spectral analysis as illustrated in figure~\ref{fig:spectra_velocity}$a$. The results show that the neural-network models successfully reproduce the energy content in the streamwise velocity component~(denoted by~$\phi_{uu}$) at different wavelengths.  However, for the wall-normal velocity fluctuations~($\phi_{vv}$) and spanwise velocity fluctuations~($\phi_{ww}$), the network models exhibit limitations in reconstructing energy at the smallest wavelengths and specifically such errors in the smallest scales increase with increasing target wall-normal position.

The power-spectral density obtained for the polymeric-shear stress~(denoted by~$\phi_{\tau_{p,\,\mathrm{xy}} \tau_{p,\,\mathrm{xy}}}$) and the trace of polymer stress~($\phi_{\mathrm{tr}(\tau_{p}) \mathrm{tr}(\tau_{p})}$) are depicted in figure~\ref{fig:spectra_velocity}$b$ for different wall-normal positions. We observe that the energetic structures correspond to wavelengths that are almost one order of magnitude smaller than those observed in the velocity fluctuations~(refer to figure~\ref{fig:spectra_velocity}$a$). This reveals that the polymer activity is predominantly concentrated in small-scale structures compared to the flow scales. Consequently, this suggests that the employed neural-network model needs to reconstruct fine-scale polymer stress fields from coarse energy-containing features in the velocity-fluctuations. %(for~E-predictions) or wall fields (for~V-E-predictions).

For the E-predictions (figure~\ref{fig:spectra_velocity}$b$), where exact velocity fields from DNS at the same wall-normal position are used as inputs, we observe the ability of the model to reconstruct the features containing energy at different wavelengths more accurately with minimal errors observed in the smallest scales. However in the case of V-E-predictions, where the wall inputs to the network feature large-scale energy-containing features~(not shown here), the performance of the network is reduced in reconstructing the energy distribution of features at smaller scales, and rather the model tends to predict the large-scale features in the polymer-stress fields. These observations underscore the importance of providing accurate velocity-fluctuation fields as inputs to the FCN. Although the auxiliary velocity fields have a very small error compared to DNS velocity fields (give a number), they lack certain information when reconstructed by FCN using wall inputs in V-E-predictions, the network is unable to capture the polymer-stress features at small and medium scales.

\subsection{Effects of small-scale velocity fluctuations on the polymer stress predictions}\label{subsec:result_filter}
Spectral analysis reveals that the energetic scales in velocity fluctuations are approximately one order of magnitude larger than those in the polymer stress quantities of interest. Consequently, it becomes challenging for the FCN to reconstruct polymer stress information at finer scales using the large energy-containing structures in velocity fluctuations. Furthermore, for E-predictions, DNS velocity fluctuation fields are used as inputs whereas for V-E-predictions, the (auxiliary) inputs are the predicted velocity fluctuations at a target wall-normal position that exhibits a good agreement with the DNS velocity fluctuation fields. It should be highlighted that the (auxiliary) velocity fluctuation fields employed in V-E-predictions quite accurately comprise the large-scale features as observed with DNS fields, although they exhibit certain differences in the small-energy containing features~(see also figure~\ref{fig:spectra_velocity}$a$). Therefore, if the large energy-containing structures in velocity fluctuations are responsible for the accurate prediction of polymer stresses, there should be minimal differences between E and V-E predictions of polymer stress components. However, we observe that small-scale polymer stress fluctuations in V-E predictions are under-predicted compared to E-predictions~(see figure~\ref{fig:stress}). This indicates that small-scale velocity fluctuations in DNS fields are crucial for the accurate prediction of polymer stresses. Thereby, the observations also signify that polymer stresses and velocity fluctuations are strongly coupled at fine, low-energetic scales.

\begin{figure}
\begin{centering}
\includegraphics[width=0.85\linewidth]{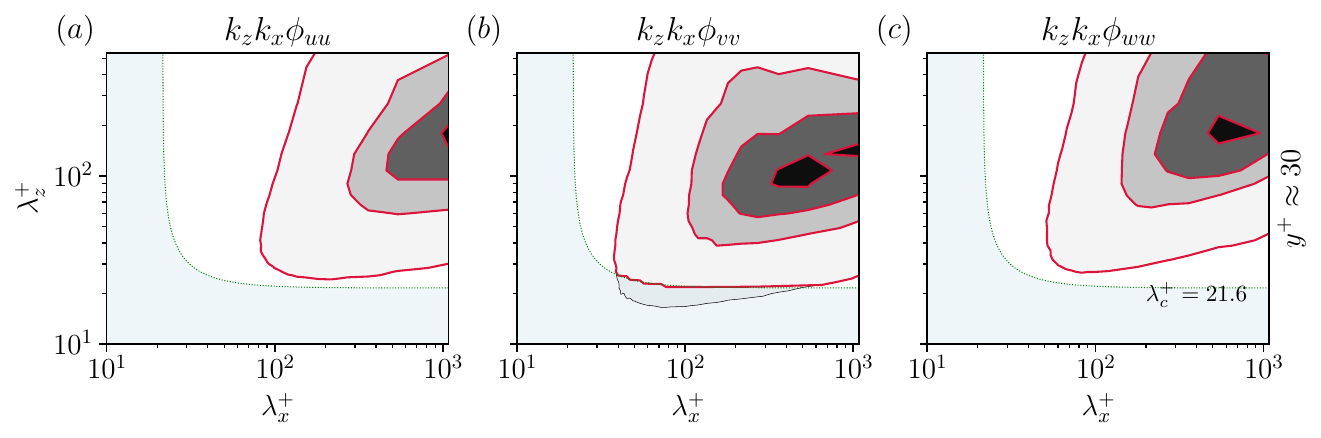}
\captionof{figure}{Pre-multiplied two-dimensional power-spectral density of~$(a)$~streamwise, $(b)$~wall-normal and~$(c)$~spanwise velocity-fluctuations at~$y^+ \approx 30$. The contour levels contain 10\%, 50\%, 80\%, 99\% of the maximum power-spectral density. Shaded contours in grey refer to DNS data, while contour lines in red indicate the corresponding energy levels after filtering with~$\lambda_c^+ = 21.6$. Filtered scales are indicated by the shaded region in blue.}
\label{fig:psd_inputs}
\end{centering}
\end{figure}

We investigate the relationship between polymer stress and velocity fluctuations at smaller scales by systematically removing the turbulent kinetic energy at these finer scales and observing the resulting polymer stress predictions by FCN. The low-pass filtering of velocity fluctuations~(as outlined in~\S\ref{subsec:filtering}) is performed to evaluate the accuracy of prediction of polymer stresses by utilizing only the large energy-containing features in velocity fluctuations. This approach not only helps in identifying the effects of small-scale features in the input velocity-fluctuation fields on the prediction of polymer stress but also enables a scale requirement recommendation for velocity fluctuations from potential experimental investigations to estimate the polymer stress information. The effect of filtering velocity-fluctuations using the pre-multiplied two-dimensional~(2D) power-spectral density~(PSD)~$k_z k_x \phi_{kk} \left(\lambda_x^+,\lambda_z^+\right)$ for the velocity-fluctuations~($k\in \{u,v,w\}$) at~$y^+ \approx 30$ as shown in figure~\ref{fig:psd_inputs}. The figure illustrates the effect of filtering the small-scale velocity-fluctuations with a wavelength threshold of~$\lambda_c^+ = 21.6$.

\begin{figure}
\begin{centering}
\includegraphics[width=0.75\linewidth]{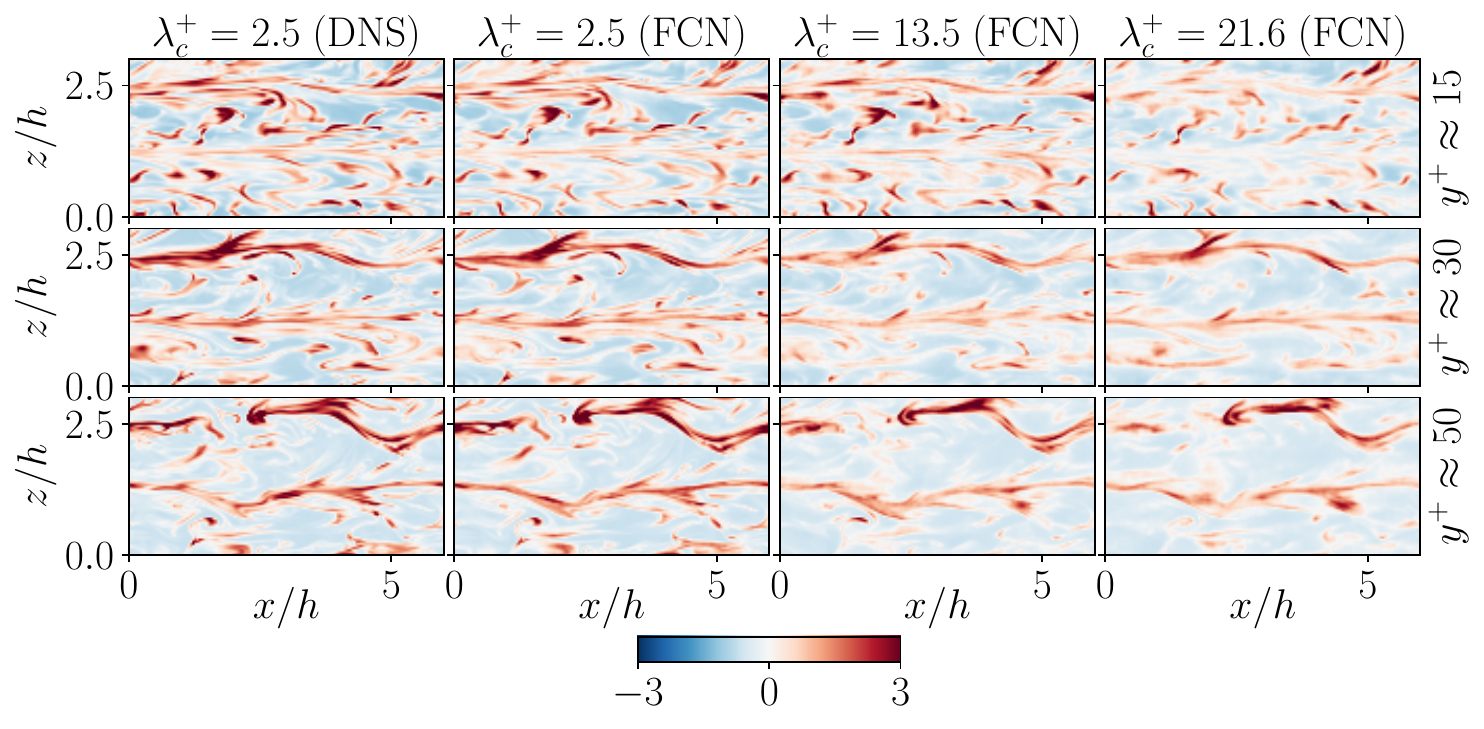}
\captionof{figure}{A sample trace of the polymer-stress fluctuation field is plotted at different-wall normal locations with corresponding predictions from FCN using inputs with different cut-off wavelengths of the velocity-fluctuation fields. The fields are scaled with the respective RMS values.}
\label{fig:qual_trace}
\end{centering}
\end{figure}

\begin{figure}
\begin{centering}
\includegraphics[width=0.9\linewidth]{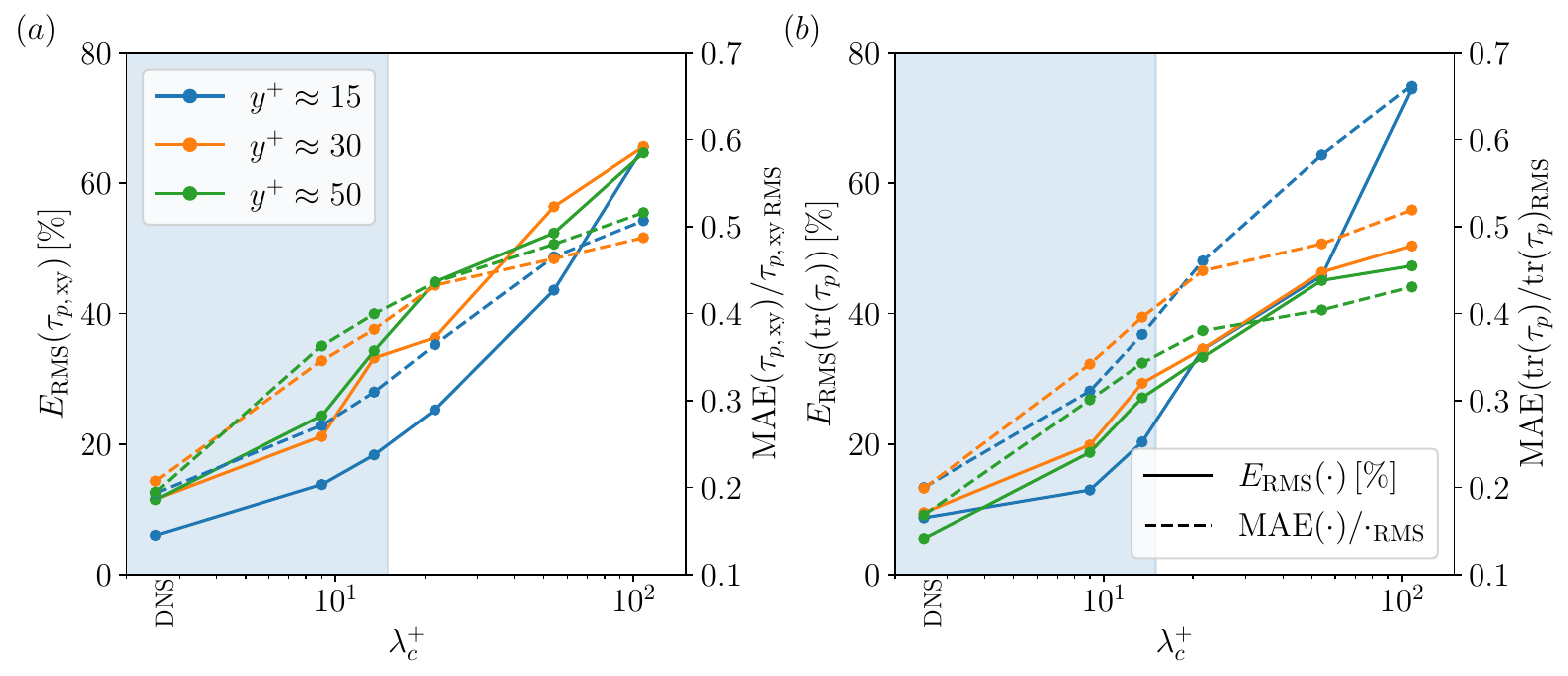}
\captionof{figure}{Variation of the errors in predicting root-mean-squared fluctuations of~$(a)$~polymer-shear stress and~$(b)$~trace of polymer stress with respect to the cut-off wavelength~$\lambda_c^+$ of velocity-fluctuation fields at respective wall-normal locations. The cut-off wavelength for DNS simulations is $\lambda_c^+ = 2.5$ based on the spanwise resolution. The corresponding normalized-mean-absolute errors are also indicated.}
\label{fig:MAE}
\end{centering}
\end{figure}

A sample instantaneous field of the trace of polymer stress in the test dataset is shown in figure~\ref{fig:qual_trace}. The reference DNS field at different wall-normal locations is shown alongside the corresponding predictions obtained from FCN, with respective inputs at different cut-off wavelengths. It is observed that the FCN predictions successfully capture all the different scales present in the reference DNS. As the cut-off wavelength in the input velocity-fluctuations to FCN increases, the predicted outputs begin to lack certain small-scale features. At~$\lambda_c^+=13.5$, where the absence of small scales contributes to a relative loss~($\Delta \mathcal{K}$) of less than $0.01\%$ of turbulent kinetic energy in the inputs at different~$y^+$ considered here, the errors in the prediction of small scale become significant, as illustrated in figure~\ref{fig:qual_trace}. Further for~$\lambda_c^+=21.6$, constituting a relative loss of less than $0.07\%$ of TKE of inputs, we observe that the small-scale features are increasingly depleted in the predictions. Nevertheless, the predictions still manage to capture the regions with large polymer extension, where~$\mathrm{tr}(\tau_p)$ is higher. A similar observation for the polymer shear-stress is provided in Appendix~\ref{app:B}.

Overall, the variation of errors in predicting the corresponding RMS quantities of the polymer stress quantities of interest is shown in figure~\ref{fig:MAE} for different values of cut-off wavelengths of inputs. The corresponding normalized mean-absolute errors are also indicated. The shaded region in the plot corresponds to the input velocity-fluctuations provided to the FCN, corresponding to a relative loss~($\Delta \mathcal{K}$) of less than $0.02\%$ of turbulent kinetic energy due to filtering. Overall, the mean errors observed at different target wall-normal locations exhibit a logarithmic growth with respect to the cut-off wavelength of the input velocity-fluctuations. This also explains the observation of higher errors in V-E predictions when using wall inputs to predict polymer stress quantities of interest at different~$y^+$ compared to E-predictions. Although the predicted velocity-fluctuations from V-predictions resemble close to that of DNS reference, there is a logarithmic increase in prediction errors of polymer stresses when utilizing the auxiliary predictions of velocity-fluctuations which lacks energy content at certain scales (see figure~\ref{fig:spectra_velocity}$a$).

In addition, for retrieval of polymer-stress information from possible near-wall experimental velocity fields using FCN, it becomes necessary to resolve finer scales, typically lower than 10 viscous lengths, to obtain more reliable polymer stress behaviour at the smallest scales. This hypothesis will be confirmed more conclusively in the next section where we further examine the effects of artificially excluding small scales from the DNS velocity fields.

\subsection{Spectral analysis on the effects of small-scale velocity fluctuations}\label{subsec:result_spec_analysis}

\begin{figure}
\begin{centering}
\includegraphics[width=0.8\linewidth]{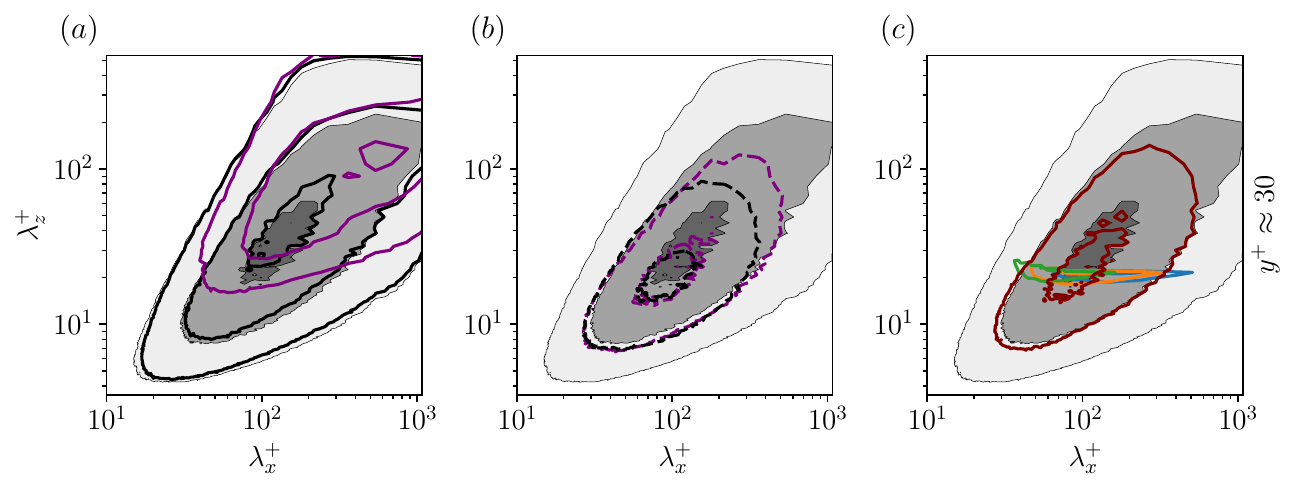}
\captionof{figure}{Pre-multiplied two-dimensional power-spectral density of trace of polymer stress at~$y^+\approx 30$. The shaded contours in all sub-figures correspond to the spectra obtained from DNS samples~($\mathrm{tr}(\tau_p)_\mathrm{DNS}$) in the test dataset.$(a)$ {\it Spectra when filtered resp. unfiltered velocity fields are used as inputs}: black contours with~DNS velocity-fluctuations as input~($\mathrm{tr}(\tau_p)_\mathrm{FCN,DNS}$), $\mathrm{purple}$ with~filtered velocity-fluctuations using a cut-off wavelength of~$\lambda_c^+=21.6$~($\mathrm{tr}(\tau_p)_\mathrm{FCN,Filt}$). $(b)$ {\it Spectra of errors}.  The spectra of the difference between reference~(DNS) and predicted polymer-stress-fluctuations using $(\mathrm{black})$~DNS velocity-fluctuations as input~($\mathrm{tr}(\tau_p)_\mathrm{DNS}$-$\mathrm{tr}(\tau_p)_\mathrm{FCN,DNS}$), $(\mathrm{purple})$~filtered velocity-fluctuations~(with $\lambda_c^+ = 21.6$) as input~($\mathrm{tr}(\tau_p)_\mathrm{DNS}$-$\mathrm{tr}(\tau_p)_\mathrm{FCN,Filt}$). $(c)$ {\it Spectra of the difference between unfiltered and filtered cases}: The spectra of the difference between~reference DNS velocity-fluctuations and the filtered velocity-fluctuations with~$\lambda_c^+ = 21.6$ for $u,v,w$ components are indicated in~blue, orange and green contour lines, respectively, while $\mathrm{brown}$~depicts the spectra of the difference between the predicted polymer-stress-fluctuations with unfiltered and filtered inputs ~($\mathrm{tr}(\tau_p)_\mathrm{FCN,DNS}$-$\mathrm{tr}(\tau_p)_\mathrm{FCN,Filt}$). 
The contour levels in all sub-figures and contour lines in $(a)$~contain~10\%, 50\% and 80\% of the maximum power-spectral density, while contour lines in~$(b,c)$ indicate 10\% and 50\% of the respective maximum power-spectral density.}
\label{fig:psd_outputs}
\end{centering}
\end{figure}

Identifying the influence of small-scale features in the velocity-fluctuation fields on the polymer stress, we probe the distribution of energy in the predicted polymer stresses due to the absence of small, low-energetic scales in the velocity fluctuations. The pre-multiplied two-dimensional power-spectral density of the trace of polymer stress ($\mathrm{tr}(\tau_p)$  at~$y^+ \approx 30$ is depicted in figure~\ref{fig:psd_outputs}. Figure ~\ref{fig:psd_outputs}$a$ shows the 10$\%$, 50 $\%$ and 80 $\%$ contours of reference~(DNS) polymer stress by shaded contours, against the FCN predictions of trace of polymer stress using DNS velocity-fluctuation fields as input to FCN~($\mathrm{tr}(\tau_p)_\mathrm{FCN,DNS}$) (black contour lines) and filtered velocity fluctuations as input to FCN ($\tau_p)_\mathrm{FCN,Filt}$) (purple contour lines).

The predictions using exact velocity fields closely capture the energy content across different scales compared to the reference~(DNS) polymer stress~(black contours are almost identical to shaded contours in figure~\ref{fig:psd_outputs}$a$). However, it should be noted that there are always some small errors in predicting the trace of polymer stress fields using DNS velocity-fluctuations as input. These errors stem from the absence of certain correlations in the inputs, such as the wall-normal gradients of velocity-fluctuations and the time history of velocity-fluctuations.

On the other hand, filtering the input velocity-fluctuations with~$\lambda_c^+ = 21.6$ so that they lack the very finest scales, results in a very significant deviation in the power density spectra of the predicted polymer stress (purple contours are very different from shaded contours). Most notably, the energy-containing structures in the predictions are at larger spatial scales than in the DNS. Additionally, the predictions now lack certain small-scale energy components. This shows that polymers interact with very small scales in the velocity fields in viscoelastic turbulence. The similarity between purple contours (predictions with filtered velocity fields) and green contours in figure~\ref{fig:spectra_velocity}$b$~(V-E predictions) confirms that errors in V-E predictions stem from the absence of the smallest scales in the auxiliary velocity fields.

The difference between the FCN predictions and the DNS reference polymer stress field represents the error in the predictions. To analyse this error quantitatively, we show the error spectra in figure~\ref{fig:psd_outputs}$b$. The error spectra illustrate the difference between the reference and predicted polymer-stress-fluctuations using DNS velocity-fluctuations as input~($\mathrm{tr}(\tau_p)_\mathrm{DNS}$-$\mathrm{tr}(\tau_p)_\mathrm{FCN,DNS}$,~black dashed contour), and the difference between reference~(DNS) and predicted polymer-stress-fluctuations using filtered velocity-fluctuations~with a cut-off wavelength of $\lambda_c^+ = 21.6$, as input~($\mathrm{tr}(\tau_p)_\mathrm{DNS}$-$\mathrm{tr}(\tau_p)_\mathrm{FCN,Filt}$)) is depicted by a purple dashed contour). From figure~\ref{fig:psd_outputs}$b$, we observe that the errors are highest at the small yet energetic scales of the trace of polymer stress, while large scales are less affected by errors. This confirms the visual finding from figure 5, that the largest scales are relatively well captured by V-E predictions.

Finally from figure~\ref{fig:psd_outputs}$c$, we examine the effect of filtering small scales in the velocity-fluctuations on the differences introduced in the predictions by the FCN. Here, the spectra of the difference between the predicted polymer-stress-fluctuations with DNS inputs~(refer to figure~\ref{fig:psd_outputs}$a$, black contour lines) and the predicted polymer-stress-fluctuations~(refer to figure~\ref{fig:psd_outputs}$a$, purple contour lines) with filtered inputs~($\mathrm{tr}(\tau_p)_\mathrm{FCN,DNS}$-$\mathrm{tr}(\tau_p)_\mathrm{FCN,Filt}$) is plotted with brown contour lines. In addition, we include the spectra of the difference between~reference DNS velocity-fluctuations and filtered velocity-fluctuations with~$\lambda_c^+ = 21.6$ for $u,v,w$ components~(respectively indicated with~blue, orange and green contour lines). The latter contours are very narrow, which shows that changes in velocity fields due to filtering mostly occur at a narrow range of scales. However, predictions of polymer stresses are affected at a strikingly large range of scales (brown contours).
Figure~\ref{fig:psd_outputs}$c$ hence shows that the loss of turbulent kinetic energy in the small-scales in the velocity-fluctuations does not necessarily correspond to the loss of energy at similar scales in the polymer stress. This observation is aligned with the studies of~\cite{nguyen2016}, where they showed from scale-to-scale analysis~\citep{casciola2007,xi2013,valente2014} of isotropic viscoelastic turbulence, that the loss of turbulent kinetic energy at a given small scale does not necessarily correspond to a gain of polymer energy at the same scale, and vice versa.

\subsection{Interpretation of predictions with filtered velocity fluctuations} \label{subsec:result_interpretation}

\begin{figure}
\begin{centering}
\includegraphics[width=0.575\linewidth]{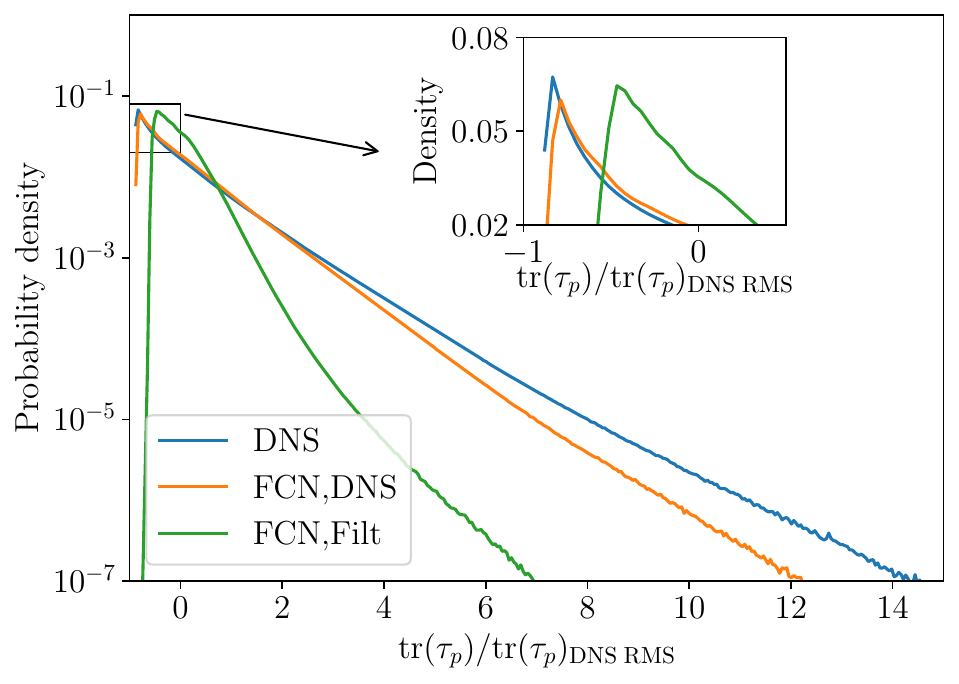}
\captionof{figure}{Probability density function of the RMS-normalized trace of polymer stress identifying the distribution of the percentage of data-points in the test dataset with a bin size of 0.045. (Inset) Magnified view of the peak of distribution.}
\label{fig:histogramtrA}
\end{centering}
\end{figure}

Overall, we observe that the small-scale features in the velocity-fluctuation fields are crucial for the estimation of polymer stress. In this section, the reason for such an observation is hypothesized based on polymer-flow interaction and obtained predictions from FCN. We start with the observation of the probability distribution function of polymer stress. The probability distribution function for the trace of polymer stress is shown in figure~\ref{fig:histogramtrA}. The predictions from FCN using DNS input velocity-fluctuations closely capture the resulting probability distribution function observed with the DNS stress fields in the test dataset. However, utilizing the filtered velocity-fluctuation fields with~$\lambda_c^+ = 21.6$ results in the distribution of the data to be closer to the mean value. The negative and positive fluctuations of the trace of the polymer stress are under-predicted indicating that the relaxation behaviour of the polymer is over-predicted and the maximum stretch of polymers is under-predicted by FCN. The over-prediction and under-prediction of polymer stress may stem from the fact that the FCN is optimized to capture the mean behaviour of the polymer-stress-fluctuations for the given velocity-fluctuation input. Because of the alteration of the input velocity-fluctuation field via filtering of small scales, the FCN is unable to predict well the extreme behaviour of polymer stress and thereby converges close to the mean behaviour and hence the resulting prediction of the standard deviation of the polymer stress quantities of interest are lower in comparison to that observed with the DNS fields. Consequently, this contributes to growing errors with respect to the cut-off wavelength of input velocity-fluctuations. However, there may be physical reasons attributed to the difficulty in predicting polymer stress without small-scale features in velocity fluctuations that are independent of the chosen technique to model the relationship between polymer stress and velocity fluctuations. Future work will involve performing direct numerical simulations with filtered velocity fluctuations to identify the effects of small-scale velocity fluctuations on the polymer stress fluctuation fields. Additional tasks also include transfer learning based approaches to evaluate the performance of the network models at different Weissenberg numbers.

\begin{figure}
\centering
\includegraphics[height=70pt]{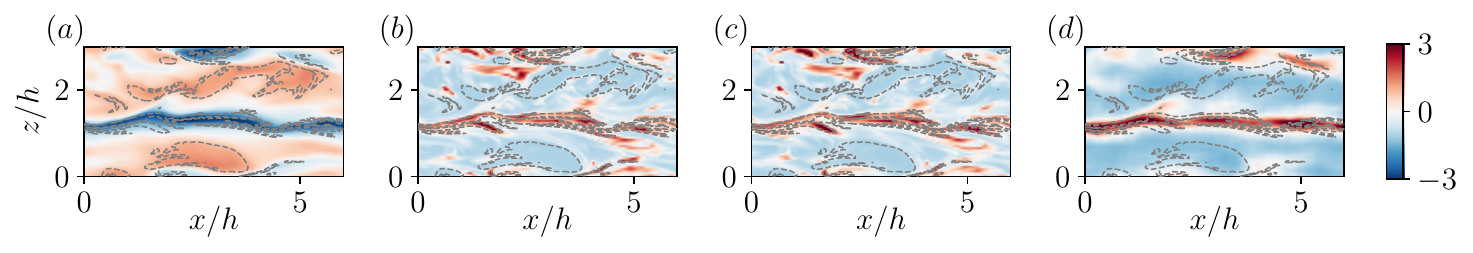}
\captionof{figure}{Comparison of RMS-scaled ($a$)~streamwise velocity-fluctuation field with ($b$)~$\mathrm{tr}(\tau_p)$ from test dataset at~$y^+ \approx 50$. The corresponding ($c$)~E-prediction and ($d$)~V-E-prediction are shown. Strong anti-correlation zones are contoured.}
\label{fig:correlation}
\end{figure}

\begin{figure}
\begin{centering}
\includegraphics[width=0.8\linewidth]{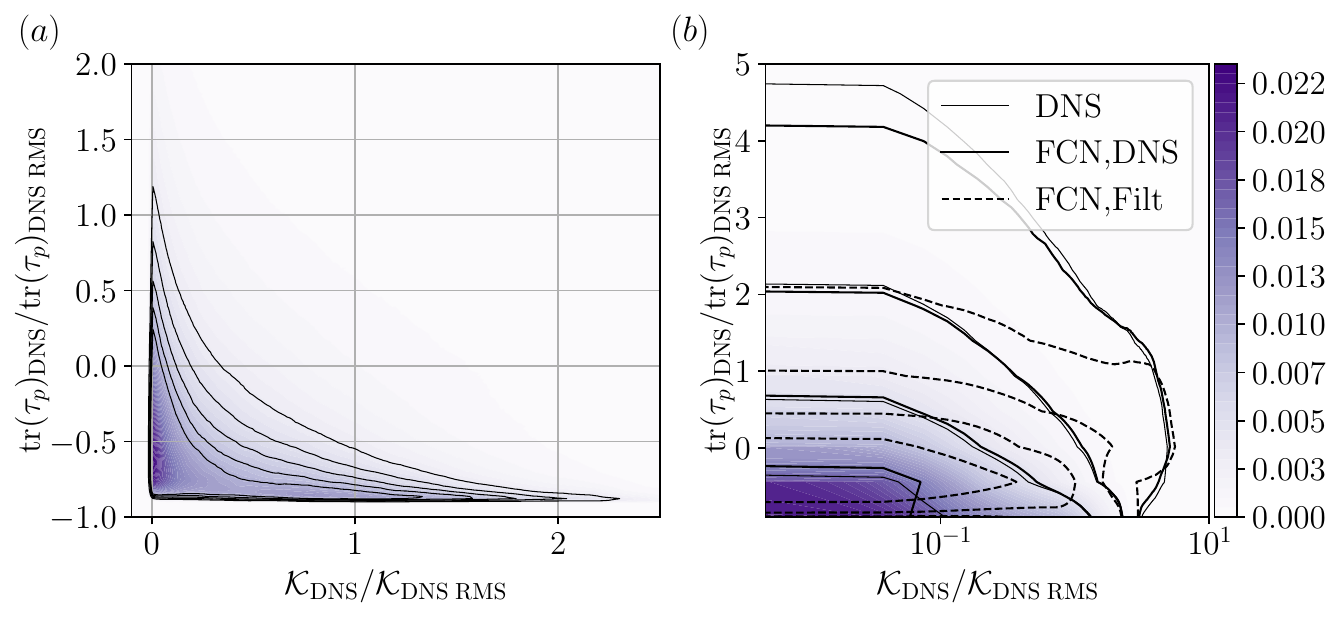}
\captionof{figure}{Normalized joint probability density function between turbulent kinetic energy and the trace of polymer stress obtained from~$(a)$~DNS samples in test dataset and~$(b)$~FCN predictions using DNS velocity-fluctuations~(FCN,DNS) and filtered velocity-fluctuations~(FCN,Filt) at~$\lambda_c^+=21.6$.}
\label{fig:jpdf_trA}
\end{centering}
\end{figure}

Filtering small scales in the velocity-fluctuation fields results in an alteration of the turbulent kinetic energy of the flow. In the buffer region that is considered in the study, there exists an anti-correlation between streamwise velocity-fluctuation and the fluctuations of the trace of polymer stress (as shown in figure~\ref{fig:correlation}), thereby an increase in polymer stresses~(and consequently polymer energy) leads to decrease in the turbulent kinetic energy of the fluid~(see also figure~\ref{fig:jpdf_trA}$a$).

The RMS-normalized joint probability distribution of the prediction errors in the trace of polymer stress with respect to turbulent kinetic energy is probed in figure~\ref{fig:jpdf_trA}$a$. From the figure, the relationship between the trace of polymer stress~(indicative of polymer stretch and elastic energy stored by polymers) and the turbulent kinetic energy of the flow from DNS samples in the test dataset is observed. It is evident that the polymers are highly stretched in regions with low turbulent kinetic energy and vice versa. This suggests that polymers extract turbulent kinetic energy from the fluid, particularly in the considered wall-normal position in the buffer region. Moreover, the relaxation of the polymer stress, which can occur in the core of the streamwise vortices, releases energy into the fluid flow~\citep{xi2019}.

From figure~\ref{fig:jpdf_trA}$b$, it is observed that the under-prediction of the trace of polymer stress fluctuations is significant in the regions of lower turbulent kinetic energy. This indicates that the effect of filtering small scales results in over-prediction of relaxation behaviour of polymer stress and under-prediction of maximum polymer elongation and thereby maximum trace of polymer stress, particularly in the regions with lower turbulent kinetic energy. Thus, due to the alteration of the distribution of small-scale input velocity-fluctuations, the FCN model is unable to reproduce the features of energy extraction by polymers from the turbulent flow. Hence, it is pertinent to capture accurately the turbulent kinetic energy in the flow and that small-scale velocity-fluctuations are required to accurately capture the extraction of turbulent kinetic energy from flow by polymers.

The results from the present study suggest that polymer stress can be estimated by directly providing potential experimental measurements of wall quantities (V-E predictions) or near-wall velocity fluctuations (E predictions) as inputs to the network model. It is important to note that experimental fields may contain noise, necessitating retraining of the network model using transfer learning methodologies to optimize the model weights with the acquired experimental dataset. However, in the current supervised learning framework, such approaches require the presence of reference experimental stress fields for retraining the network models, which is not applicable in this case. It was recently shown that Lagrangian stretching fields extracted from particle image velocimetry (PIV) serve as indicators of polymer elongation fields under certain conditions~\citep{kumar2023}, but their applicability to stress field substitutes in channel flow turbulence remains to be investigated. Therefore, we propose that as an initial step to gain insight into elastic stresses from experimental measurements, interpolation (for matching resolution) and de-noising techniques~\citep{discetti_noise,nekkanti_noise,yousif_noise} can be employed to retrieve accurate wall fields or velocity fluctuations from experimental data. These refined velocity inputs can then be coupled with the proposed DNS-trained network models to obtain predictions of instantaneous elastic stress fields within the probed domain. 

The present work can also be extended towards identifying coherent structures in the flow and thereby employing SHapley Additive exPlanations (SHAP) algorithm to explain the importance of the coherent structure~\citep{cremades24} which can improve the understanding of the dynamical role of coherent structures in viscoelastic turbulent flow. Hence the present work serves as a starting point, holding potential for many future works to understand better the polymer physics in drag reduced flows. 
\section{Conclusions}

The present work highlights the capability of a data-driven approach to perform non-intrusive sensing in viscoelastic turbulent flows. Here we demonstrate the ability of CNN-based models to accurately reconstruct the velocity fluctuations in viscoelastic turbulence close to the wall, utilizing the two wall-shear fluctuation components and the wall-pressure fluctuations as inputs. Additionally, the network models successfully reproduce the polymeric-stress fluctuation fields from the DNS velocity-fluctuation fields. Moreover, the feasibility of these network models to extract polymer-stress-fluctuation fields of interest solely from wall input fluctuations and predicted velocity fluctuations is explored. Overall, the network effectively reconstructs the large-scale features of the polymer-stress fields using wall inputs and predicted velocity fields. Furthermore, the developed models exhibit enhanced accuracy in predicting quantities of interest during the hibernation intervals, facilitating a deeper understanding of the underlying physics during low-drag events when the model is deployed in a practical application. These non-intrusive-sensing models hold valuable applications in experimental settings~\citep{vinuesa23}, enabling the determination of polymeric stresses in turbulent flows from velocity fields or wall inputs, which otherwise would be challenging or impossible to quantify experimentally. However, accurately predicting the polymer stresses from wall inputs is found to be more challenging than predicting velocity fields from the same, since wall inputs typically do not contain fingerprints of the smallest scales, and we show that small scales in velocity fields are connected to a wide range of scales in polymeric stress fields.

Aimed towards extracting polymer-stress information from velocity-fluctuation fields in a possible experimental investigation of viscoelastic turbulent channel flow, a number of FCN models are trained with inputs corresponding to different thresholds of small-scale wavelengths in the velocity-fluctuation fields. We find that accurately capturing the turbulent kinetic energy in the near-wall fields is crucial for retrieving second-order turbulent statistics of polymer stresses. Specifically, experimental acquisition of velocity fluctuations would in principle need to resolve the finer scales, smaller than 10 viscous units, to obtain more reliable polymer stress behaviour at the smallest scales, which determines the accurate rate of energy transfer from flow to polymers. We show that this in turn facilitates an accurate estimation of the RMS of polymer stresses in the buffer region of viscoelastic turbulent channel flow.

Concluding, the results demonstrate the potential of data-driven models to predict instantaneous fields in non-Newtonian wall turbulence, which is useful for flow control or estimation of stress fluctuations that cannot be measured. As a next step, de-noised experimentally measured velocity fields and wall pressure measurements could be given as inputs to the network that has been trained by simulations. Furthermore, direct numerical
simulations with filtered velocity fluctuations can be performed to further isolate the physical mechanisms behind how small-scale velocity fluctuations influence the polymer stress fluctuation fields at a wide range of scales.
\appendix
\section{Performance metrics for network models} \label{app:A}

In the present study, hyperparameter tuning was conducted to investigate the sensitivity of various parameters, including network model capacity, depth, kernel size, batch size, and learning rate, on the errors in predicting the RMS of velocity and polymer stress fluctuations. Approximately 15 different training runs were performed to optimize these parameters, and a more thorough exploration of the hyperparameter space could further enhance prediction performance. In the present study, a fully convolutional neural network containing $3\times 3$ kernels and 31 hidden layers constituting $985,105$ trainable parameters is employed~(refer figure~\ref{fig:1}). The initial weights for the network are distributed randomly with a Gaussian distribution and a scheduled learning rate ($\alpha$) was provided to the Adam algorithm~\citep{kingma} in the form $\alpha = \alpha_0 a^{\mathrm{epoch}/b}$, with $\alpha_0$ corresponding to initial learning rate and $a,b$ are the tunable parameters denoted by learning rate drop and epoch drop, respectively (here, epoch corresponds to one entire pass of training data). In this study, $\alpha_0 = 0.001$, $a = 0.5$ and $b=40$ and a batch size of 16 was employed in the training of FCN.

Additionally, U-Net~(with five skip connections) and Generative Adversarial Network~(GAN) models~(with about 2 million trainable parameters in the generator) were also trained, and corresponding evaluations revealed comparable performance in terms of accuracy to that achieved with the FCN in the present study. Hence, the FCN with the optimal training parameters as obtained with the exploratory study is employed in the present work.

The prediction errors for V-predictions are provided in table~\ref{tab:table1} and various metrics for E-predictions and V-E-predictions are detailed in table~\ref{tab:table2}. The error metrics reported in table~\ref{tab:table1} and table~\ref{tab:table2} are obtained from the samples contained in the test dataset and averaged over three different training runs for the FCN model. The variances are indicated in the braces.

\begin{table}
\begin{center}
\begin{tabular}{ccccc}
& & & \multicolumn{1}{c}{${i}$} & \\ \cline{3-5}
Parameters       & $y^+$    & ${u}$                    & ${v}$                      &${w}$\\
\midrule
 & 15 & 0.14 ($\pm$ 0.01) & 0.32 ($\pm$ 0.11) & 0.41 ($\pm$ 0.02) \\
$\left<\mathrm{MAE}(i)\right>_t/i_\mathrm{DNS\,RMS}$ & 30 & 0.29 ($\pm$ 0.01) & 0.44 ($\pm$ 0.06) & 0.53 ($\pm$ 0.02) \\
 & 50 & 0.47 ($\pm$ 0.01) & 0.79 ($\pm$ 0.15) & 0.59 ($\pm$ 0.04) \\ [2pt]

 & 15 & 2.62 ($\pm$ 1.2) & 13.11 ($\pm$ 2.7) & 8.02 ($\pm$ 3.8) \\
$E_\mathrm{RMS}(i)$ & 30 & 5.21 ($\pm$ 0.3) & 15.90 ($\pm$ 4.5) & 10.62 ($\pm$ 1.9) \\
 & 50 & 12.73 ($\pm$ 3.9) & 24.72 ($\pm$ 6.5) & 14.72 ($\pm$ 4.4) \\ [2pt]

 & 15 & 0.996 ($\pm$ 0.003) & 0.889 ($\pm$ 0.040) & 0.932 ($\pm$ 0.018) \\
$R(i)$ & 30 & 0.942 ($\pm$ 0.002) & 0.763 ($\pm$ 0.015) & 0.771 ($\pm$ 0.010) \\
 & 50 & 0.811 ($\pm$ 0.001) & 0.623 ($\pm$ 0.011) & 0.643 ($\pm$ 0.010) \\ [2pt]
\end{tabular}
\caption{\label{tab:table1} Model-averaged errors in V-predictions}
\end{center}
\end{table}

\begin{table}
\begin{center}
\begin{tabular}{cccccc}
& & & \multicolumn{1}{c}{${i}$} & & \\ \cline{3-6}
Parameters       & $y^+$    & ${\tau_{p\;\mathrm{xy}}}\;[E]$  & ${\mathrm{tr}(\tau_p)}\;[E]$                      &${\tau_{p\;\mathrm{xy}}}\;[V-E]$ & ${\mathrm{tr}(\tau_p)}\;[V-E]$\\
\midrule
 & 15 & 0.19 ($\pm$ 0.01) & 0.20 ($\pm$ 0.01) & 0.45 ($\pm$ 0.01) & 0.63 ($\pm$ 0.01) \\
$\left<\mathrm{MAE}(i)\right>_t/i_\mathrm{DNS\,RMS}$ & 30 & 0.21 ($\pm$ 0.01) & 0.20 ($\pm$ 0.02) & 0.49 ($\pm$ 0.02) & 0.60 ($\pm$ 0.01)  \\
 & 50 & 0.19 ($\pm$ 0.01) & 0.17 ($\pm$ 0.02) & 0.48 ($\pm$ 0.01) & 0.55 ($\pm$ 0.01) \\ [2pt]

 & 15 & 6.03 ($\pm$ 1.8) & 8.68 ($\pm$ 1.4) & 42.52 ($\pm$ 1.0) & 54.71 ($\pm$ 2.6)\\
$E_\mathrm{RMS}(i)$ & 30 & 11.54 ($\pm$ 2.7) & 9.47 ($\pm$ 1.8) & 49.40 ($\pm$ 3.4) & 51.14 ($\pm$ 0.7)\\
 & 50 & 11.46 ($\pm$ 1.4) & 5.52 ($\pm$ 1.4) & 62.37 ($\pm$ 3.2) & 49.54 ($\pm$ 4.8)\\ [2pt]

 & 15 & 0.906 ($\pm$ 0.004) & 0.924 ($\pm$ 0.007) & 0.484 ($\pm$ 0.006) & 0.362 ($\pm$ 0.013) \\
$R(i)$ & 30 & 0.906 ($\pm$ 0.006) & 0.917 ($\pm$ 0.010) & 0.296 ($\pm$ 0.013) & 0.433 ($\pm$ 0.008) \\
 & 50 & 0.905 ($\pm$ 0.007) & 0.924 ($\pm$ 0.011) & 0.207 ($\pm$ 0.012) & 0.426 ($\pm$ 0.015) \\ [2pt]
\end{tabular}
\caption{\label{tab:table2} Model-averaged errors in E-predictions and V-E-predictions}
\end{center}
\end{table}

\section{Effects of low-pass filtering the velocity fluctuations on prediction of polymer shear stress}\label{app:B}
\begin{figure}
\begin{centering}
\includegraphics[width=1\linewidth]{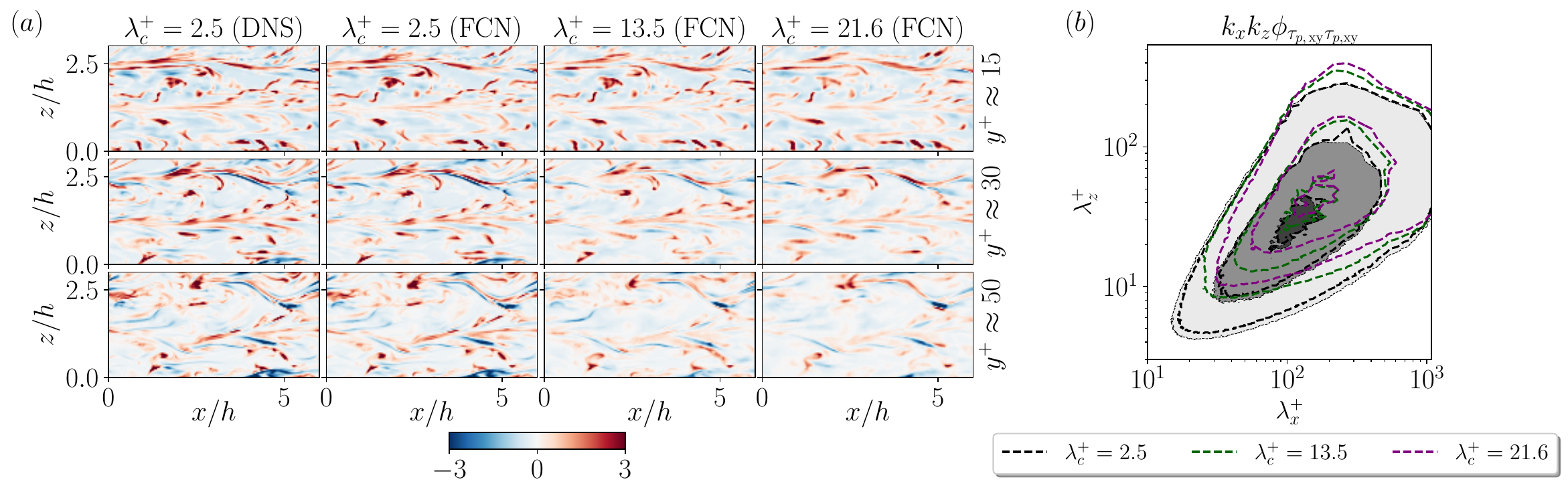}
\caption{\label{fig:qual_shear} $(a)$~A sample polymer-shear stress fluctuation field is plotted at different-wall normal locations with corresponding predictions from FCN using inputs with different cut-off wavelength of the velocity-fluctuation fields. The fields are scaled with the respective RMS values. $(b)$~Pre-multiplied two-dimensional power-spectral density of polymer-shear stress at~$y^+ \approx 30$. The contour levels contain 10\%, 50\% and 80\% of the maximum power-spectral density. Shaded contours refer to DNS data, while contour lines indicate the cut-off wavelength in the input velocity-fluctuations provided to FCN.}
\end{centering}
\end{figure}

A sample instantaneous field of the polymer shear stress~($\tau_{p,\,\mathrm{xy}}$) in the test dataset is shown in figure~\ref{fig:qual_shear}$a$. The sampled DNS field at different wall-normal locations serves as a reference while the corresponding predictions obtained from FCN with respective inputs at different cut-off wavelengths are depicted. The figure shows that the resulting predictions from FCN lack certain small-scale features with increasing~$\lambda_c^+$ for different wall-normal locations, a fact that is attributed to the absence of fine-scale features in the input fields.

Further, the distribution of energy in different scales for both the DNS and the FCN-predicted polymer-shear stress field at~$y^+ \approx 30$ is illustrated in figure~\ref{fig:qual_shear}$b$, which also identifies that the resulting predictions from FCN increasingly feature energy-containing large-scale structures and lack energy content in small scales with increasing~$\lambda_c^+$.

\section{Joint probability density function of polymer shear stress} \label{app:C}

\begin{figure}
\begin{centering}
\includegraphics[width=0.85\linewidth]{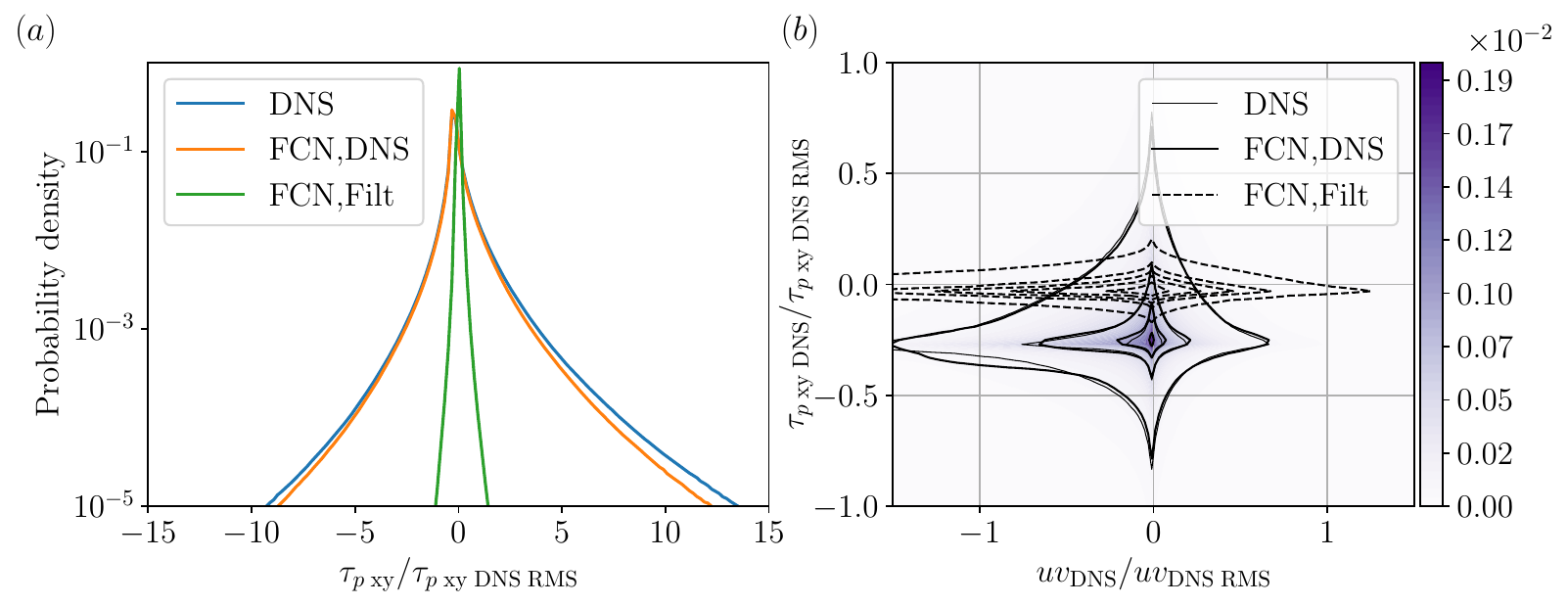}
\captionof{figure}{$(a)$~Probability density function of the RMS-normalized polymer shear stress identifying the distribution of percentage of data-points in the test dataset with a bin size of 0.18. $(b)$~Normalized joint probability density function between turbulent shear stress and polymer shear stress obtained from DNS samples in test dataset and FCN predictions using DNS velocity-fluctuations~(FCN,DNS) and filtered velocity-fluctuations~(FCN,Filt) with~$\lambda_c^+=21.6$.}
\label{fig:jpdf_shear}
\end{centering}
\end{figure}

A similar observation can be made for the prediction of the polymer shear stress with respect to turbulent shear stress as outlined in~\S\ref{subsec:result_interpretation} for the trace of polymer stress. The probability density function for the RMS-normalized polymer shear stress is shown in figure~\ref{fig:jpdf_shear}$a$. The predictions from FCN using DNS input velocity-fluctuations closely capture the probability density function observed with the DNS polymer shear stress fields in the test dataset. However, utilizing the filtered velocity-fluctuation fields with~$\lambda_c^+ = 21.6$ results in the distribution of the data to be closer to the mean value, indicating an under-prediction of the fluctuations of the polymer shear stress.

From figure~\ref{fig:jpdf_shear}$b$, observing the distribution from DNS samples, we find the polymer shear stress to mitigate the production of turbulent shear stress. Furthermore, predictions of polymer shear stress using filtered fields are under-predicted, especially in regions with negligible turbulent shear stress. Filtering the input fields alters the Reynolds shear stress, leading to increased errors in the FCN's predictions of polymer shear stress.

\backsection[Acknowledgements]{The authors acknowledge the National Academic Infrastructure for Supercomputing in Sweden (NAISS) for providing the computational resources to carry out the numerical simulations and training of convolutional network models.}

\backsection[Funding]{This work is supported by the funding provided by the European Research Council grant no.~"2021-StG-852529, MUCUS" and the Swedish Research Council through grant No 2021-04820. RV acknowledges the ERC grant no.~"2021-CoG-101043998, DEEPCONTROL".}

\backsection[Declaration of interests]{The authors report no conflict of interest.}

\backsection[Data availability statement]{The data that support the findings of this study will be openly available on GitHub--\href{https://github.com/orgs/KTH-Complex-fluids-group}{KTH-Complex-fluids-group} upon publication.}

\bibliographystyle{jfm}
\bibliography{jfm}

\end{document}